\documentclass[5p,times]{elsarticle}

\usepackage{lineno}
\usepackage[draft]{hyperref}
\modulolinenumbers[5]

\usepackage[T1]{fontenc}
\usepackage[utf8]{inputenc}
\usepackage{amsmath}
\usepackage{natbib}
\usepackage{graphicx}
\usepackage{gensymb}
\usepackage{url}
\usepackage{mathtools}
\usepackage{textcomp}
\usepackage[super]{nth}
\usepackage{chemformula}

\usepackage{esvect}

\usepackage{siunitx}
\biboptions{authoryear}

\begin{document}

\begin{frontmatter}

\title{High precision meteor observations with the Canadian Automated Meteor Observatory \\
\large Data reduction pipeline and application to meteoroid mechanical strength measurements}

\author[uwopa]{Denis Vida}
\ead{dvida@uwo.ca}
\author[uwopa]{Peter G. Brown}
\author[uwopa]{Margaret Campbell-Brown}
\author[uh]{Robert J. Weryk}
\author[bern]{Gunter Stober}
\author[nrl]{John P. McCormack}

%\address[uwoes]{Department of Earth Sciences, University of Western Ontario, London, Ontario, N6A 5B7, Canada}
\address[uwopa]{Department of Physics and Astronomy, University of Western Ontario, London, Ontario, N6A 3K7, Canada}
\address[uh]{Institute for Astronomy, University of Hawaii, Honolulu HI, 96822, USA}
\address[bern]{Institute of Applied Physics, University of Bern, Sidlerstrasse 5, CH-3012 Bern, Switzerland}
\address[nrl]{US Naval Research Laboratory, 4555 Overlook Avenue SW, Washington, DC 20375, United States}

\begin{abstract}

\textit{Context.} The mirror tracking system of the Canadian Automated Meteor Observatory (CAMO) can track meteors in real time, providing an effective angular resolution of 1 arc second and a temporal resolution of 100 frames per second.

\textit{Aims.} We describe the upgraded hardware and give details of the data calibration and reduction pipeline. We investigate the influence of meteor morphology on radiant and velocity measurement precision, and use direct observations of meteoroid fragmentation to constrain their compressive strengths.

\textit{Methods.} On July 21, 2017, CAMO observed a $\sim 4$ second meteor on a JFC orbit. It had a shallow entry angle (\ang{\sim 8}) and 12 fragments were visible in the narrow-field video. The event was manually reduced and the exact moment of fragmentation was determined. The aerodynamic ram pressure at the moment of fragmentation was used as a proxy for compressive strength, and strengths of an additional 19 fragmenting meteoroids were measured in the same way. The uncertainty in the atmosphere mass density was estimated to be $\pm 25\%$ using NAVGEM-HA data.

\textit{Results.} We find that meteor trajectory accuracy significantly depends on meteor morphology. The CAMO radiant and initial velocity precision for non-fragmenting meteors with short wakes is $\sim 0.5'$ and \SI{1}{\metre \per \second}, while that for meteors with fragments or long wakes is similar to non-tracking, moderate field of view optical systems ($\sim 5'$, \SI{\sim 50}{\metre \per \second}). Measured compressive strengths of 20 fragmenting meteoroids (with less precise radiants due to their morphology) was in the range of \SIrange{1}{4}{\kilo \pascal}, which is in excellent accord with Rosetta in-situ measurements of 67P. Fragmentation type and strength do not appear to be dependent on orbit. The mass index of the 12 fragments in the July 21 meteoroid was very high ($s = 2.8$), indicating possible progressive fragmentation.

\end{abstract}

\end{frontmatter}

% \linenumbers

\section{Introduction}

Most currently operational optical meteor observation systems consist of a fixed low-light camera operating at typical video frame rates of 25 to 30 frames per second. Such systems vary from all-sky to moderate fields of view, with plate scales at best 1 arcmin/pixel \citep[e.g.][]{jenniskens2011cams, toth2015all}. Since the appearance of low-light CCD sensors in the early 1990's, video meteor cameras have proved to be an invaluable source of data used for meteoroid orbital studies, models of meteoroid fragmentation and physical structure, as well as recovery of meteorites produced by meteorite-dropping fireballs \citep{koten2019meteors}.

Dynamical and photometric data from such cameras have allowed the densities of meteoroids, a critical parameter used in spacecraft risk models \citep{mcnamara2005meteoroid, kikwaya2011bulk}, to be systematically estimated. However, such fixed optical systems are not able to observe individual meteor fragments as they move relative to the fixed sensor during frame integration, and the pixel scale on the order of 10s of meters per pixel at \SI{100}{\kilo \metre} does not allow the meteor morphology to be fully resolved \citep{stokan2013optical}. This obscures the true amount and nature of fragmentation and deceleration, which are essential for constraining the physical properties of meteoroids through ablation modelling \citep{vojacek2019properties} and are the limiting factor in improving accuracy of meteoroid orbits \citep{vida2018modelling}. Finally, moderate field of view optical systems, in ideal conditions, can achieve meteor trajectory radiant accuracy of \ang{0.1}, which is near the limit of resolving the true radiant dispersion of the tightest (youngest) meteor showers \citep{vida2019meteorresults}.

\begin{figure*}
  \includegraphics[width=\linewidth]{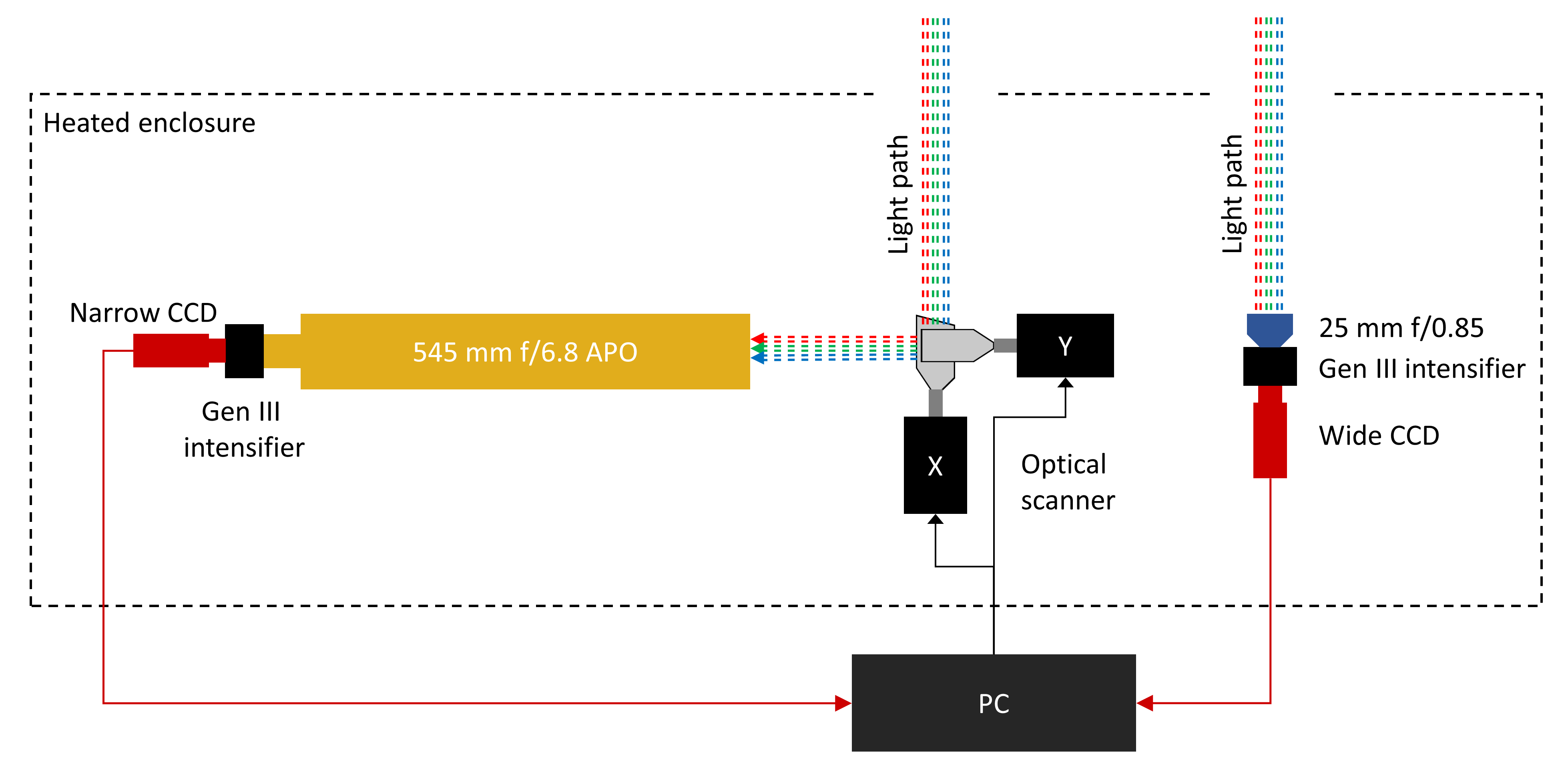}
  \caption{Block diagram of the CAMO mirror tracking system.}
  \label{fig:camo_diagram}
\end{figure*}

\begin{figure}
  \includegraphics[width=\linewidth]{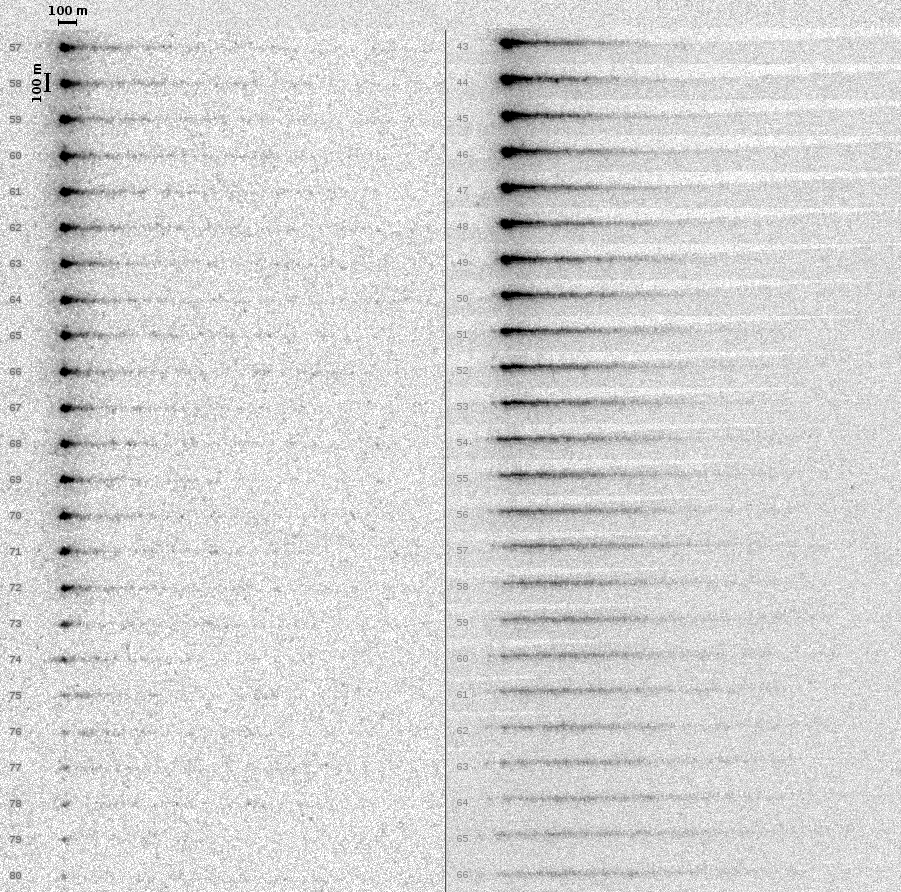}
  \caption{Composite grey-inverted image of two meteors recorded using the narrow-field camera. Both meteors are rotated and their leading edges aligned so the evolving meteor morphology is highlighted in this vertical stack, with time increasing from top to bottom and frame number given to the left. Left: An $\alpha$ Capricornid recorded on August 4, 2019 at 06:02:42 UTC. Right: An $\alpha$ Capricornid recorded on July 27, 2019 at 04:21:22 UTC. Note that even though they are of the same origin, the meteor on the left erodes away leaving a single distinct fragment with no wake, while the one on the right completely disintegrates into a long cylinder of dust.}
  \label{fig:narrowfield_meteor_stacks}
\end{figure}

The Canadian Automated Meteor Observatory's (CAMO) mirror tracking system is an optical system consisting of a wide-field camera ($\ang{34} \times \ang{34}$) which runs a real time meteor detection algorithm. Upon detection, it cues a pair of mirrors to track the meteor and redirect its light through an \SI{80}{\milli \metre} telescope with a very narrow field of view ($\ang{1.5} \times \ang{1.5}$) equipped with a 3$^{rd}$ generation image-intensifier lens coupled to a high frame rate machine vision CCD camera, giving a plate scale of 6 arcsec/px \citep{weryk2013camo}. A block diagram of the tracking system is shown in Figure \ref{fig:camo_diagram} and more details about the hardware are given in Section \ref{sec:camo_hardware}. The data collected with CAMO provide a direct way of studying the details of fragmentation for mm-sized meteoroids and offer the prospect of order of magnitude more precise meteoroid orbits as compared to classical fixed-camera systems \citep{vida2019meteorresults}. Observing meteor morphology and distinguishing the wake reveals the underlying physics of ablation and fragmentation, and allows direct measurements of individual fragments: their dynamics, differential deceleration, mass, mass distribution, and strengths. Figure \ref{fig:narrowfield_meteor_stacks} shows an example composite image of two meteors observed with the CAMO narrow field of view camera, and in supplementary materials we provide several meteor videos. Although CAMO was a unique system at the time of its construction, we note that a similar tracking system has recently been deployed at the Ond\v{r}ejov observatory in the Czech Republic. Nevertheless, in contrast to CAMO, the scientific focus the Fireball Intelligent Positioning System (FIPS) is on observing details of fragmentation of bright fireballs \citep{shrbeny2020fireball, borovivcka2020two}.

\subsection{Previous research done using CAMO data}

The CAMO mirror tracking system has been operational since 2009. An early study by \cite{subasinghe2016physical} showed that 90\% of the mm-sized meteors observed by CAMO display observable fragmentation while \cite{campbell2017modelling} recognized that light curves of meteors with short or invisible wakes (resembling single-body meteors) cannot be explained without including continuous fragmentation in their ablation model. In an effort to explain meteors with double-peaked light curves, \cite{subasinghe2019properties} found that even allowing for large compositional differences within meteoroids to produce multiple peaks, fragmentation still had to be included to explain the observed span of ablation heights. These conclusions from direct observations are in accordance with classical works \citep[e.g.][]{verniani1969structure}, and strongly support the notion that meteors cannot be modelled as single bodies and that fragmentation is an essential process of the meteor phenomenon.

\cite{stokan2013optical} directly measured optical trail widths of meteors using CAMO, a parameter vital for computing meteoroid masses from radar observations. \cite{stokan2014transverse} were the first to recognize a class of fragmenting mm-sized meteors producing large lateral fragment separations. The transverse speeds of fragments produced by these faint meteors approached \SI{100}{\metre \per \second}, constraining the range of meteoroid strengths by assuming rotational- or charge-based fragmentation. \cite{subasinghe2016physical} classified meteors by their fragmentation morphology and found that the frequency and type of fragmentation does not correlate with orbital classes, i.e. meteoroids on asteroidal-type orbits fragment in the same way and as often as those on cometary-type orbits.  \cite{subasinghe2019properties} found that meteors with double-peaked light curves on asteroidal-type orbits usually have a very sharp second lightcurve peak, indicating that the meteoroid either disrupted or suddenly released many small grains; in contrast, meteoroids on cometary-like orbits had smooth two-peaked light curves.

Using the fact that the CAMO mirror tracking system is able to gather data allowing for precise measurement of meteoroid deceleration, \cite{subasinghe2017luminous} and \cite{subasinghe2018luminous} derived luminous efficiencies of meteors by comparing the dynamic and photometric masses. Their results were broadly consistent with both theory and previous measurements, but measurement errors were still too high to accurately identify the best-matching theoretical model of luminous efficiency. In contrast to the luminous efficiencies estimated for fireballs \citep{ceplecha2005fragmentation}, \cite{subasinghe2018luminous} found a negative linear correlation of luminous efficiency with initial meteoroid mass, a trend also recently reported by \cite{capek2019small} who computed luminous efficiencies for iron meteoroids by comparing observed and simulated meteor re-radiation energies.

As meteor wakes are easily visible with CAMO, they provide an additional constraint on meteoroid ablation and fragmentation models beyond observed light curves and deceleration. \cite{campbell2013high} compared fits of the thermal disruption model of \cite{campbell2004model} and the thermal erosion model of \cite{borovicka2007atmospheric} for several CAMO meteors. In most cases, they were able to match the observed light curve and deceleration, but not the wake. \cite{campbell2017modelling} was able to fit all three features of one meteor with a short and faint wake using a modified ablation model, where short bursts of fragments were continuously released, a fragmentation model similar to the \cite{borovicka2007atmospheric} erosion model. 

This emphasizes the complexity of fragmentation for mm-sized meteoroids as well as the different modes of such fragmentation. In some instances, for example, discrete fragments are visible in CAMO imagery and the fragment release heights are measurable. For such cases, it may be possible to extend CAMO observations to infer the compressive strengths of small meteoroids, an otherwise difficult to measure parameter.

\subsection{Introduction to meteoroid strength measurements} \label{subsec:strengths_intro}

Measuring mechanical strengths of meteoroids provides insights into the corresponding surface strengths of their parent bodies. \cite{biele2009putative} thoroughly reviewed past meteoroid strength studies for the purpose of designing the landing gear and landing procedure for Rosetta's ill-fated Philae comet lander. As comets consist of the most pristine material leftover from the formation of the Solar System \citep{blum2017evidence}, understanding mechanical properties of their constituent particles informs models of dust aggregate growth in protoplanetary disks \citep{guttler2009physics}. Finally, understanding the comet mechanical surface strength is essential for models of comet activity and dust mass distribution \citep{gundlach2018tensile}.

Global-scale strengths of comets can be investigated though some types of comet and asteroid break-ups, such as rotational \citep[e.g.][]{lisse1999nucleus, davidsson2001tidal, sanchez2014strength}, or tidally-driven \citep{asphaug1996size}. At millimeter to meter scales,  strengths can be derived either theoretically \citep[e.g.][]{guttler2009physics} or from direct measurements. The measurements done in-situ are rather limited in number due to the complexity of the task. 

\cite{hornung2016first} analyzed the dust from comet 67P /Churyumov-Gerasimenko at sizes between \SI{10}{\micro \metre} and \SI{300}{\micro \metre}, collected by the COSIMA instrument onboard the Rosetta spacecraft. They found that particles smaller than \SI{100}{\micro \metre} remained largely undamaged when they collided with the instrument collection plate at velocities of several meters per second, but that particles larger than \SI{100}{\micro \metre} (i.e. optical meteor sizes) mostly fragmented upon collision into smaller grains of several tens of microns in size. They estimate that the mechanical strength of larger particles is of the order of several kilopascals ($\pm$ factor of 2), a result supported by subsequent experimental work of \cite{gundlach2018tensile}. 

Microscopic imaging of dust particles collected by COSIMA \citep{hornung2016first} shows that the dust has an agglomerate structure, with the constituent particles on the order of a few tens of microns. Their critical observation with implications for meteoroids is, quoting from \cite{hornung2016first}: ``... these sub-units, which we denoted as `elements' for not-fragmented dust particles are essentially within the same size range as the individual `fragments' dispersed by the impact-fragmented dust particles. This observation seems to show that the fragments are not formed by the impact, but pre-existing in the parent dust agglomerate, and simply broken apart during the impact.'' In section \ref{subsec:july21_even_morphology} we discuss a meteor observed by CAMO which supports this statement. We also note that the sizes of elemental grains observed by \cite{hornung2016first} match the grain size distribution derived from meteor flares \citep{simonenko1968separation} and meteoroid erosion models \citep{borovicka2007atmospheric, vojacek2019properties}.

Due to the unsuccessful landing of the Philae lander, during which it bounced off the surface of 67P/~ Churyumov-Gerasimenko several times, \cite{biele2015landing} were able to estimate that the compressive strength of softer surface regions was on the order of \SI{1}{\kilo \pascal} on 10 cm-to-meter scales, the maximum being between \SIrange{2}{3}{\kilo \pascal}. The lander finally landed on a hard surface with a compressive strength of \SI{>2}{\mega \pascal}.

For the purpose of understanding the physics of planetesimal formation in protoplanetary disks, many authors have performed experiments in which they investigate growth of dust agglomerates from micron-sized dust through sticking \citep{blum2000growth, blum2000experiments, krause2004growth}. The growth to mm-sized fractal dust aggregates is supported by the theoretical relative collision velocities found in protoplanetary disks, but as the particles grow in size so do the velocities, which should in theory lead to fragmentation of these larger particles \citep{blum2008growth}. To overcome the mm-size fragmentation boundary, lower collision velocities are needed to facilitate further sticking. It is believed that gas compression may play a role at those sizes \citep{kataoka2013fluffy, birnstiel2016dust}. To determine how low the velocities must be, \cite{weidling2012free} performed a microgravity experiment where they investigated the sticking of mm-sized \ch{SiO2} dust aggregate analogs at velocities as low as \SI{1}{\milli \metre \per \second}. They reproduced the sticking behaviour and shown that the contact strength of dust aggregates in their experiments is at least \SI{640}{\pascal}. Most recently, \cite{kimura2020tensile} developed a novel analytical model of strengths of dust particles depending on their porosity and volume - their model showed good agreement with in-situ cometary dust and in-atmosphere meteoroid strength measurements.

\subsection{Strength measurements from meteoroid fragmentation in the atmosphere} \label{subsec:strengths_atm}

Mechanical strengths can also be derived from fragmentation of meteoroids in the atmosphere. If the height of fragmentation is known, one can assume that the dynamic ram pressure exerted on the meteoroid at the moment of fragmentation is a proxy for the compressive strength. Note that other authors may call this the ``tensile'' strength \citep{baldwin1971ablation, trigo2006strength} due to the assumption of e.g. differential thermal heating causing internal thermal stress, or bending due to a non-spherical shape. Because we do not use a thermal model, nor attempt to measure the meteoroid shape, we assume that the fragmentation happens due to mechanical failure caused the by pressure difference between the front and the back of the meteoroid, thus we use the term ``compressive'' strength \citep{kataoka2017dust}. Nevertheless, we note that thermal stress due to differential thermal heating may play a role for \SI{>1}{\milli \metre} sized meteoroids \citep[it is negligible at smaller sizes;][]{verniani1969structure}. Past models of thermal stresses within meteoroids during entry often assume cm-sized meteoroids to be non-porous (basalt-like) stony particles. These are expected to show large internal temperature gradients \citep{elford1999thermally, bariselli2020aerothermodynamic} which should in theory catastrophically fragment before ablation begins \citep{jones1966effects}. However, high resolution CAMO observations reported by \cite{subasinghe2016physical} show that gross fragmentation only occurs in 5\% of the cases, and most happen after the onset of ablation. It is unclear whether thermal fragmentation remains a plausible process for highly porous dust aggregates; the recent work by \cite{markkanen2019scattering} shows that \SI{0.5}{\milli \metre} porous aggregates are able to withstand internal temperature gradients of more than \SI{150}{\kelvin}. This mechanism should be investigated in more detail. Here we assume dynamic pressure is the dominant process of gross fragmentation. 

Using a model of atmosphere mass density, the dynamic pressure can be simply computed using the following expression:

\begin{equation} \label{eq:dyn_press}
    P_{dyn} = \Gamma v^2 \rho(h)
\end{equation}

\noindent where $\Gamma$ is the drag coefficient (usually assumed to be unity in free molecular flow, appropriate to most events observed by CAMO), $v$ is the meteoroid speed at the moment of fragmentation, and $\rho(h)$ is the atmosphere mass density at the height of fragmentation $h$.

It has long been understood that if smaller meteoroids do fragment when they enter the atmosphere, this fragmentation occurs at the dynamic pressure of around \SIrange{1}{2}{\kilo \pascal} \citep{verniani1969structure}. Past in-atmosphere meteoroid strength studies either used a rule of thumb or an ablation model to estimate when this fragmentation may occur;  \cite{blum2014comets} gives an overview of past work. \cite{trigo2006strength} measured strengths in the range from \SI{400}{\pascal} for the Draconids to \SI{340}{\kilo \pascal} for the Taurids, assuming that meteoroids disrupted at the point of maximum brightness. On the other hand, \cite{borovicka2007atmospheric} fit observed meteor light curves and decelerations to a meteoroid ablation model and concluded that except for bright fireballs, fragmentation does not coincide with the point of maximum brightness. They estimated that the compressive strengths of more compact parts of Draconid meteoroids are in the range of \SIrange{5}{20}{\kilo \pascal}. They also note that meteoroid erosion (continuous fragmentation into constituent grains) can start earlier and may not be due to mechanical forces, and that only fireball flares are caused by disruption (catastrophic fragmentation) which could reveal the compressive strength. Regardless of the possible mechanisms of disruption/fragmentation, precise determination of the fragmentation time can set concrete upper limits to mechanical bulk strengths of meteoroids. 

\cite{borovivcka2020two} analyzed compressive strengths of meteorite-dropping fireballs and found that although only the strongest parts of meteoroids (\SIrange{20}{40}{\mega \pascal}) survive the atmospheric flight, the fragmentation starts at strengths as low as \SIrange{40}{120}{\kilo \pascal}. They suggest that these weak parts of meteoroids are reassembled and cemented debris of asteroid collisions. \cite{shrbeny2020fireball} investigated strengths of smaller non-meteorite dropping fireballs and found that they start to release grains at strengths between $4$ and \SI{62}{\kilo \pascal}. In both these two papers, the authors used a meteoroid ablation and fragmentation model to fit the observed light curve and the deceleration. They assumed that the wake-producing erosion into $\mathrm{\mu m}$-sized grains began due to mechanical failure. This approach is different from \cite{borovicka2007atmospheric} who suggested that the onset of erosion might be due to thermal effects (e.g. a temperature gradient near the surface of the body), as the whole meteoroid doesn't experience mechanical failure at once, but only the grains get released from the surface at a sustained rate.

\subsection{Motivation and overview}

This paper was preceded by two theoretical papers which investigated the limits of meteor trajectory accuracy achievable with CAMO. When computing meteoroid orbits, understanding the accuracy of two separate components is essential: the meteoroid pre-atmosphere velocity and the radiant. In \cite{vida2018modelling} an ablation model was used to simulate meteoroids of different physical properties as they would be observed by various optical observation systems, including CAMO. In particular, they investigated whether measuring the meteor velocity at the beginning of the luminous trajectory produced accurate pre-atmosphere velocity measurements. They found that meteoroids producing optical meteors can significantly decelerate before their ablation becomes visible, up to \SI{750}{\metre \per \second}, which is an order of magnitude more than the velocity measurement precision of the CAMO system. \cite{vida2018modelling} have also shown that this deceleration is highly influenced by meteoroid density and other physical properties, implying that ablation models must be used to fit the observed meteor to invert for the true pre-atmosphere velocity; thus the ultimate limitation on velocity accuracy is the efficacy of the adopted ablation model.

\cite{vida2019meteortheory} and \cite{vida2019meteorresults} investigated the radiant accuracy that can be achieved by CAMO and found that it is an order of magnitude more accurate ($\ang{\sim 0.01}$) than what is needed to measure the model-estimated true physical radiant dispersion of the most compact meteor showers, specifically the Draconids. They also found that existing methods of meteor trajectory estimation were not suitable for the high-precision CAMO data, so an improved method was developed which simultaneously uses both the geometrical and dynamical information to constrain meteor trajectory solutions, but without forcing a kinematic model.

In this work, we first describe recent  upgrades to the CAMO hardware, software, and data reduction procedure in detail. In section \ref{sec:reduced_meteors} we present the first results of high-precision meteor reductions on CAMO data and discuss the implications of the observed meteor morphology on trajectory accuracy. Next, in section \ref{sec:strengths_measurements}, we estimate compressive strengths of select meteors by taking the dynamic pressure at the moment of observable gross fragmentation as a proxy of strength. The measurements were only done for meteoroids where the exact moment of fragmentation was directly visible in the CAMO recordings. In section \ref{subsec:strengths_sensitivity_analysis} we perform a brief sensitivity analysis of computing dynamic pressures in practice, with a special focus on the uncertainty of the atmosphere density. Next, in section \ref{subsec:july21_event} we describe an event observed with the CAMO tracking system on July 21, 2017 which shows the exact moment of gross fragmentation that prompted us to develop a method of direct compressive strength measurement \citep{vida2018canadian}. Finally, in section \ref{subsec:strength_survey} we present compressive strength results for a small number of events by applying our approach to several meteors showing gross fragmentaion.

\section{CAMO mirror tracking system specifications}

The first version of the CAMO system started regular operations in 2009 and was described by \cite{weryk2013camo}. CAMO is comprised of two identical systems in Southwestern Ontario, Canada, separated by \SI{\sim 45}{\kilo \metre}. The first is located near Tavistock and is co-located with the Canadian Meteor Orbit Radar (CMOR) (\ang{43.26420} N, \ang{80.77209} W, \SI{329}{\metre}), while the other is at the Elginfield Observatory (\ang{43.19279} N, \ang{81.31565} W, \SI{324}{\metre}). Both systems are pointed roughly northward at an elevation of \ang{45} to avoid sunlight, moonlight, and the galactic plane. Their common volume overlap is optimized for heights between \SI{70}{\kilo \metre} and \SI{120}{\kilo \metre}. This configuration limits the maximum convergence angle between stations to \ang{\sim 27}, but this has no detrimental effect on meteor trajectory accuracy due to the fine astrometric scale of the data when an appropriate trajectory solver is used \citep{vida2019meteorresults}.

\subsection{System hardware} \label{sec:camo_hardware}

In mid-2017, the system hardware was upgraded to extend CAMO's operational lifespan. Both the wide and the narrow-field cameras, which are lens coupled to Generation 3 image intensifiers, were replaced with 14-bit Prosilica GX1050 digital progressive scan CCD cameras which use a Gigabit Ethernet interface, and have an image resolution of $1024\times1024$ pixels. The video is cropped to $900\times900$ pixels, as the edges of the field of view are not covered by the intensifier. The wide-field camera is operated at 80 frames per second, and the narrow-field camera at 100. The intensifiers were upgraded to new \SI{18}{\milli \metre} diameter ITT FS9910 series Generation 3 image intensifiers with 64 line pairs per millimeter resolution, providing a close to 1:1 match to the camera resolution. The intensifier on the wide-field camera is operated continuously during observations, but the narrow-field intensifier is gated and only turns on if a meteor is being tracked. This saves up to 99\% of intensifier time, significantly prolonging its lifetime. 

The lens setup remained the same as before, with a \SI{25}{\milli \metre} f/0.85 lens on the wide-field, giving a field of view of $\ang{34}\times\ang{34}$. The narrow-field optics are also unchanged and consist of an \SI{80}{\milli \metre} aperture APO telescope with a \SI{545}{\milli \metre} focal length. As the telescope is looking at mirrors with an effective radius of \SI{50}{\milli \metre}, the narrow-field setup's focal ratio is reduced to f/11, giving it a field of view of $\ang{1.5}\times\ang{1.5}$ and a plate scale of 6 arcsec per pixel. The effective meteor limiting magnitude of the wide-field system is about $+4.5^M$, and the narrow-field $+7^M$.

\begin{figure}
  \includegraphics[width=\linewidth]{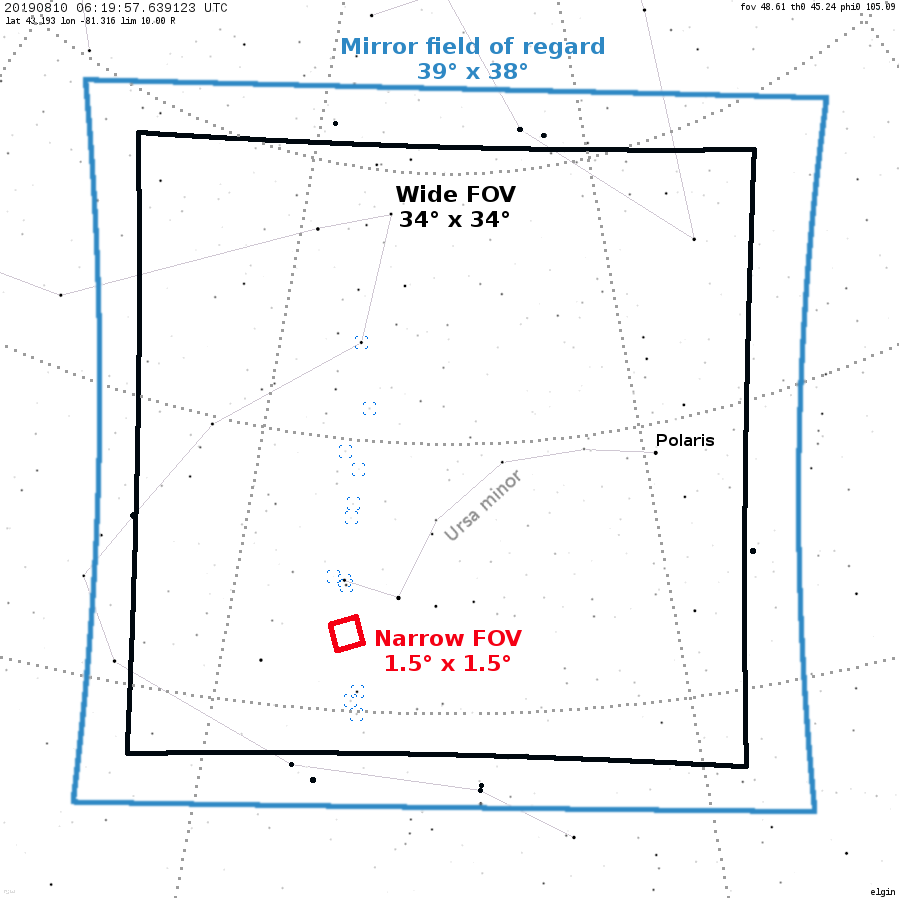}
  \caption{Fields of view of all CAMO optical subsystems at the Elginfield site. The narrow field of view can move to any location inside the mirror field of regard.}
  \label{fig:camo_fovs}
\end{figure}

Assuming that ideal centroiding can improve position measurements by a factor of three, this system is at the limit of the average atmospheric seeing in Southwestern Ontario, thus no further improvement in resolution can be achieved under these conditions. The two mirrors on orthogonal axes are attached to a Cambridge Technology 6900 optical scanner, a galvanometer-based system with a maximum slew rate of 2000 deg/s and a field of regard of $\ang{39} \times \ang{38}$. Figure \ref{fig:camo_fovs} shows the comparison of the fields of view of all optical subsystems. The mirrors on the optical scanner are precisely positioned by changing the voltage of each axis between \SI{-10}{\volt} and \SI{10}{\volt} using a 16-bit digital-to-analog converter, giving an angular step-size resolution of $\sim 2.2$ arcsec/ADU, equivalent to 1/3 of a pixel in the narrow-field camera.

The clock on the computers that operate each camera is GPS conditioned using the network time protocol (NTP), but the video frames are timestamped on the camera to avoid timing errors due to Ethernet network latency. We found that the camera's internal clock drifts over time relative to the NTP computer clock, so we apply a frame time correction by occasionally checking the temporal drift and fitting a linear model through the time differences during nighttime observations. Total time drifts remain sub-frame between time calibrations, which occur every two hours during operations. 

\begin{figure}
  \includegraphics[width=\linewidth]{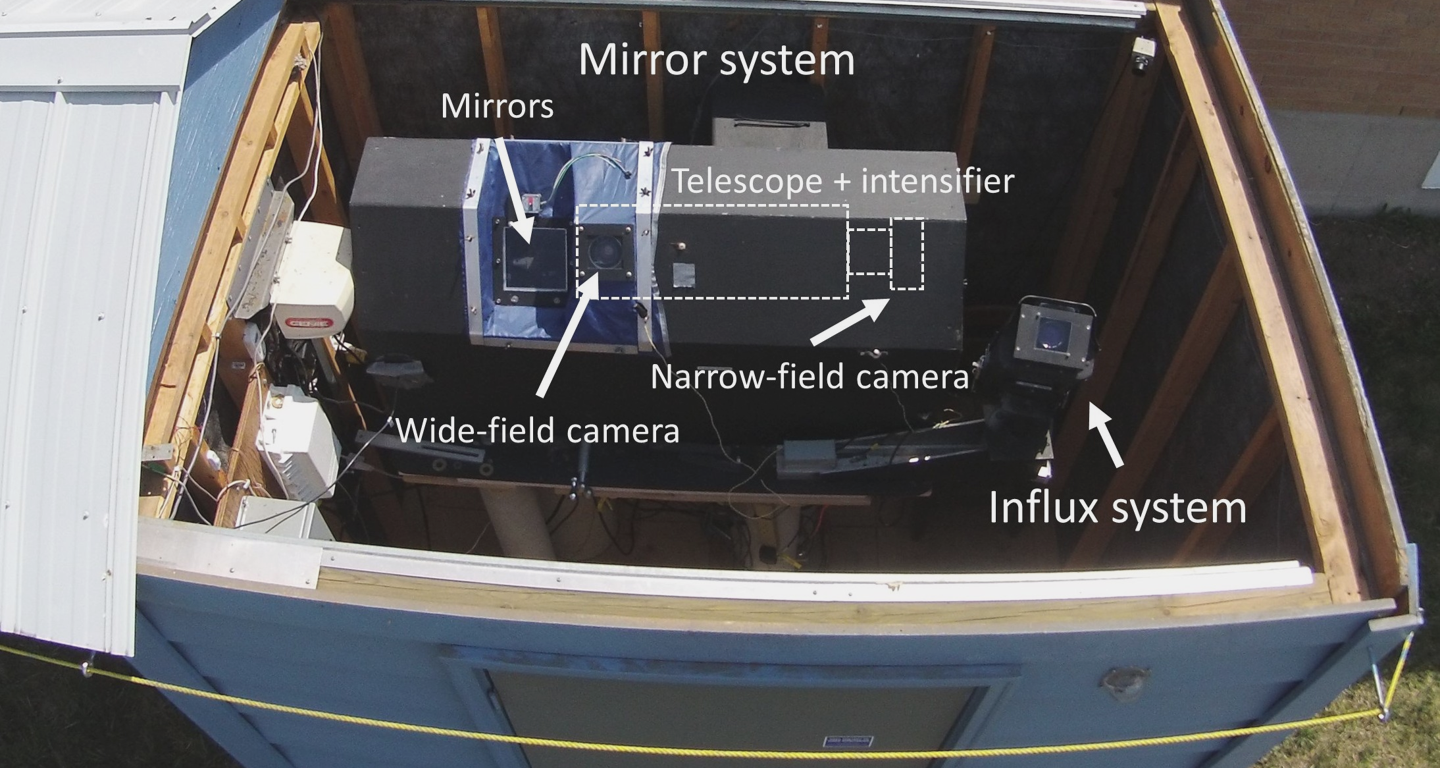}
  \caption{Layout of CAMO systems. The enclosure of the mirror tracking system with BK7 glass windows of individual cameras is shown. The position of the telescope (inside the block enclosure) is outlined and superimposed on the image. The CAMO influx camera \citep{Musci2012} is at the lower right, but is not used in this work.}
  \label{fig:camo_layout}
\end{figure}

The optical system is contained within a weather-resistant enclosure inside a roll-off roof shed which only opens during optimal weather conditions \citep{weryk2013camo}. The system layout is shown in Figure \ref{fig:camo_layout}.

\subsection{Detection software and tracking}

The meteor detection algorithm used by CAMO has been described in detail in \cite{weryk2013camo}. In this paper we only give a short summary, but thoroughly describe the calibration methods which were not discussed in detail in \cite{weryk2013camo}.

As image-intensified video is dominated by high frequency shot noise, we use a normalised first-order low-pass finite impulse response filter in our detection algorithm \citep{weryk2012simultaneous}. This approach eliminates bright and short bursts of noise, while being sensitive to any medium frequency events (such as meteors) which appear above the static background. The shot noise does not typically have enough trigger pixels to form a detection. The algorithm runs in real time on the wide-field video feed and once it detects a meteor in 8 frames, it fits a constant angular speed model based on these detections. The mirrors then slew to and track according to the model-predicted motion, having their positions updated 2000 times a second. Due to the mirror inertia and high speed of position updates, their motion becomes fluid, allowing the imaging to match the reference frame of each tracked meteor. A record of mirror position at every update is kept, making high-precision astrometry using narrow-field data possible. Note that there are no encoders which read the actual position, thus the ``commanded to" and the actual position may differ. We find that this simple tracking algorithm is able to keep meteors within the narrow field of view camera in most cases over their full visible trajectory. Finally, note that because of the tracking delay of 8 frames (\SI{0.1}{\second} at 80 FPS), the high-precision position measurements are also delayed, which may cause initial velocities of meteors to be underestimated if computed using narrow-field data.

\section{Calibration}

\subsection{Operational plates for tracking}

To steer the mirrors when a meteor is being tracked, wide-field camera imagery coordinates ($w_x, w_y$) are converted into analog-digital units ($h_x, h_y$) of the 16-bit voltage controller which positions the mirrors. As the telescope optical axis is fixed with respect to the mirror,  it also points at $h_x, h_y$ - this always corresponds to the centre of the narrow-field camera image axis. This mapping is achieved using a \texttt{guide} plate, an affine transform mapping between $w_x, w_y$ and $h_x, h_y$. Thus when a good \texttt{guide} plate is used, tracked meteors should be in the centre of the field of view of the narrow-field camera. \texttt{guide} plates are created by pairing stars in the wide-field camera imagery with the same stars centred in the narrow-field camera, and fitting an affine \texttt{AFF} type plate (see \ref{appendix:aff_plate} for details). As the narrow-field camera has a very small field of view, it would be difficult to do this pairing manually, so a mosaic of narrow-field images taken across the whole mirror field of regard is constructed to produce a first fit when the optical system components are installed or have been moved. 

The fitting procedure requires the paired stars to be centered in the narrow field of view images to produce a quality fit. When the mosaic is created, the stars can be anywhere inside the narrow field of view. To ``virtually'' center them, a \texttt{scale} plate is used which maps offsets from the narrow-field image centre ($\Delta n_x, \Delta n_y$) to offsets in mirror units ($\Delta h_x, \Delta h_y$). Thus the \texttt{scale} plate is used to compute $h_x, h_y$ coordinates of paired narrow-field stars, and a \texttt{guide} plate can be fit.

\begin{figure}
  \includegraphics[width=\linewidth]{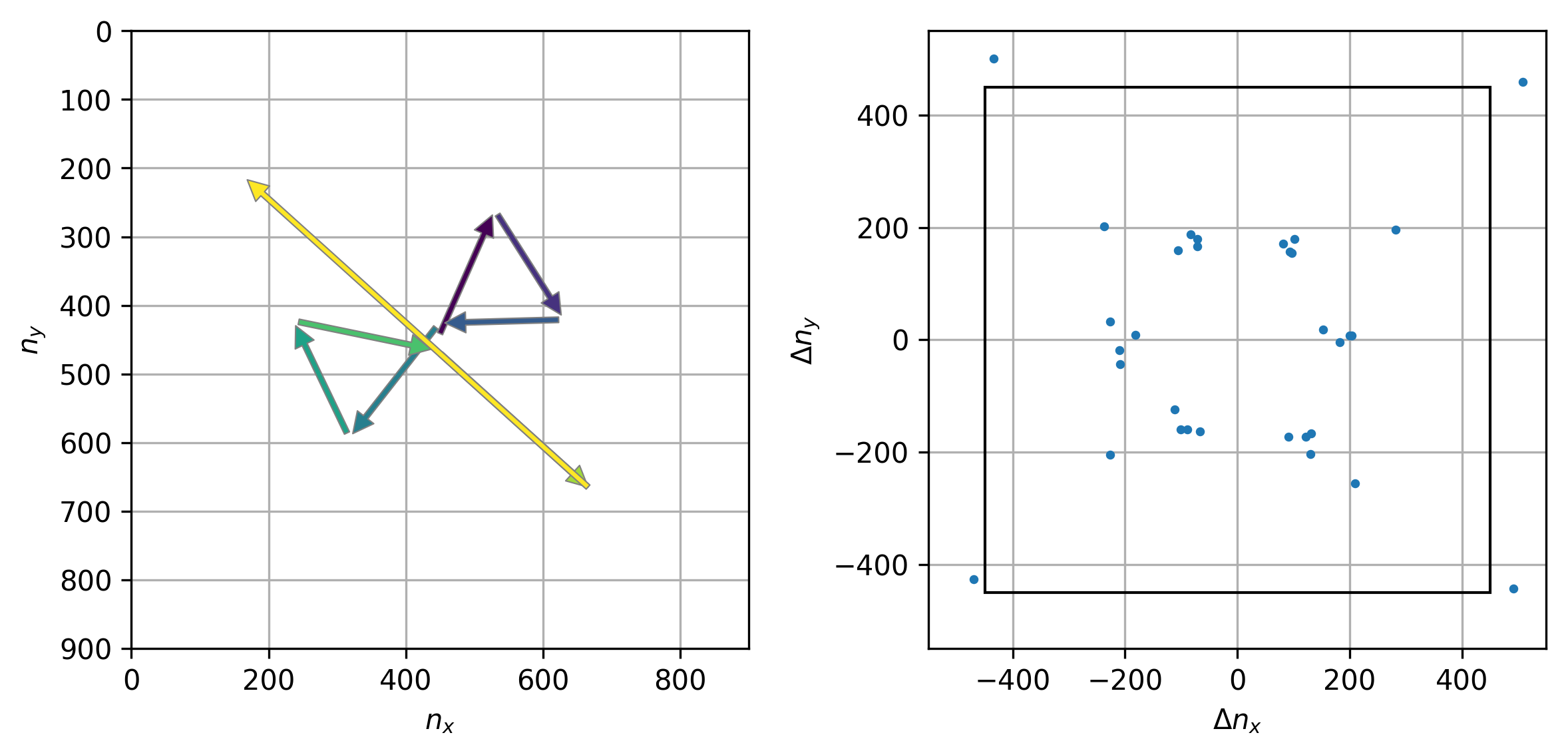}
  \caption{Left: One of four patterns used for moving a star within the narrow-field system which produces a hexagonal pattern in the \texttt{scale} plate phase space. The pattern starts after the star has been centered. The first movement is represented with the darkest arrow. Right: Points organized in a hexagon in the \texttt{scale} plate phase space produced by all four patterns of movement. The black rectangle outlines the range of possible input values of the scale plate, which is $\pm$ half the image size in each dimension. The points outside the rectangle are used to improve the fit at the edges of the phase space and were obtained by moving a star past the centre of the image. One such movement is represented by the lightest long diagonal arrow in the left inset, which produced a point around $\Delta n_x = 450$, $\Delta n_y = -450$ in this example.}
  \label{fig:scale_phase_space_hexagon}
\end{figure}

A \texttt{scale} plate is made by locking and centering the mirrors onto a bright star, then moving the mirrors by small steps in a specific pattern to obtain pairs of $\Delta n_x, \Delta n_y$ and $\Delta h_x, \Delta h_y$. An affine type (AFF) plate is fit on those data pairs. Figure \ref{fig:scale_phase_space_hexagon} shows the movement pattern which produces a set of points in the \texttt{scale} plate parameter space which form a hexagon. The data points are distributed in such a way as to equalize the distance between the edges of the parameter space and the points themselves, optimally populating the parameter space. 

In general the \texttt{scale} plate does not change over long periods as it reflects the stability of the fixed effective focal length of the narrow field optics. The \texttt{guide} plate is also relatively invariant as long as the wide field and narrow field systems remain fixed relative to one another; this typically does not require updating more than a few times per year.

\subsection{Astrometry calibration plates}

Table \ref{tab:plates} summarizes all calibration plates used by CAMO. To compute meteor trajectories after data collection, image and mirror coordinates have to be converted into celestial coordinates. The astrometric calibration of the wide-field camera is done by manually pairing image stars with catalog stars and fitting an astrometric \texttt{AST} type plate (see appendix \ref{appendix:ast_plate} for details). Star positions and magnitudes are taken from the SKY2000 catalog \citep{weryk2013camo}. This procedure produces a \texttt{calib} plate which maps wide-field image coordinates $w_x, w_y$ into local horizontal celestial coordinates: the zenith angle $\theta$, the azimuth $\varphi$ measured North of East. The \texttt{calib} plate may change slightly under different thermal conditions as the wide field camera mounting moves, but this is usually a small change night-to-night (of order a few hundredths of a degree).

The \texttt{exact} plate maps mirror coordinates $h_x, h_y$ into $\theta, \varphi$ and is used for computing high-precision astrometry using narrow-field video data. The \texttt{exact} plate is created in several steps. First, a list of stars in the wide field of view sorted by their brightness is produced using the star catalog and the \texttt{calib} plate. Next, the mirrors are pointed to the 80 brightest stars using the \texttt{guide} plate. Off-center stars are moved to the center of the narrow field of view using the \texttt{scale} plate. This procedure is done every two hours during system operations to ensure the quality of the astrometry. Finally, the collected pairs of mirror coordinates $h_x, h_x$ and star coordinates $\theta, \varphi$ are used to fit an \texttt{exact} AST type plate. The \texttt{exact} plate avoids using the \texttt{calib} plate which is limited by the spatial resolution of the wide-field camera. The \texttt{exact} plate is the most time-varying, as very slight changes in the mirror directions produced by thermal effects may cause drift in the encoder positions relative to the sky. The \texttt{exact} plate has to be computed and updated nightly, sometimes even multiple times per night to maintain the full narrow-field positional accuracy.

\begin{table*}[t] %floating table at the top of the page
	\caption{Meaning of the various plates used for CAMO calibration. $w_x$, $w_y$ are pixel coordinates of the wide-field camera, $n_x$, $n_y$ are pixel coordinates of the narrow-field camera, and $h_x$, $h_y$ are mirror encoder coordinates. $\theta$, $\varphi$ are local horizontal coordinates, namely the zenith angle and the azimuth (+N of due E), as described in \ref{appendix:ast_plate}.}
	\label{tab:plates} % is used to refer this table in the text
	\centering % used for centering table
	
	\begin{tabular}{l c c c l} % centered columns (5 columns)
	\hline\hline % inserts double horizontal lines
	
	Plate & Input & Output & Type & Description \\ % table heading
	\hline % inserts single horizontal line
	
	\texttt{calib} & $w_x$, $w_y$               & $\theta$, $\varphi$        & AST & Wide-field astrometric calibration. \\
	\texttt{guide} & $w_x$, $w_y$               & $h_x$, $h_y$               & AFF & Pointing mirrors to the given wide-filed pixel. \\
	\texttt{scale} & $\Delta n_x$, $\Delta n_y$ & $\Delta h_x$, $\Delta h_y$ & AFF & Narrow-field offsets from image centre to \\
	      &                            &                            &     & offsets in mirror encoder coordinates. \\
	\texttt{exact} & $h_x$, $h_y$               & $\theta$, $\varphi$        & AST & Mirror astrometric calibration. \\
	\hline %inserts single line
	\end{tabular}
\end{table*}

\section{Data reduction}

\subsection{CAMO tracking system weblog}

\begin{figure}
  \includegraphics[width=\linewidth]{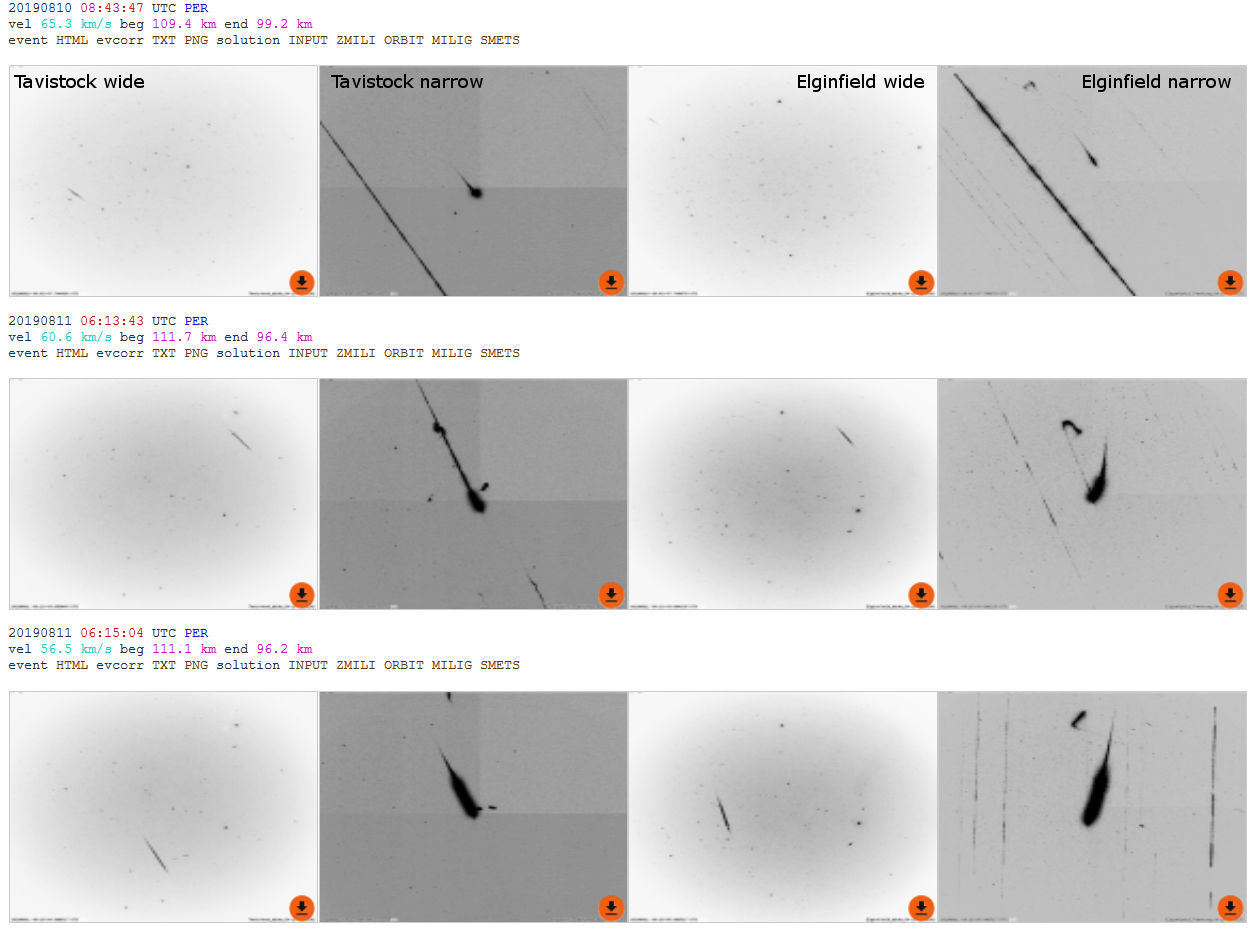}
  \caption{Screenshot of the CAMO weblog page showing three Perseids. Co-added video frames from all sites and both wide and narrow field of view cameras are shown.}
  \label{fig:camo_weblog}
\end{figure}

After a full night of automated meteor detection is complete, event data  (cutouts of wide- and narrow-field videos) are sent to a central server where meteoroid orbits are computed based on the operational astrometric plates. Every morning, a weblog page is generated with images and preliminary orbits of events based on wide-field imagery automatically detected the previous night. CAMO successfully detects and tracks about a dozen meteors to a limiting stellar magnitude of $+4^M$ at two stations on an average clear night with nominal meteor activity. Videos of tracked events can be inspected which helps to identify events suited for further in-depth study (through manual data reduction). Figure \ref{fig:camo_weblog} shows a screenshot of the weblog page showing three Perseid meteors from August 13, 2019.

\subsection{Manual reduction of wide-field data}

The reduction of data from the wide FOV camera is described in detail in \cite{weryk2013camo}. Briefly, the ASGARD automated meteor detection software \citep{Weryk2007} stores raw video frames of meteor detections. Flat fields are created by median co-adding a large number of video frames from throughout the night, which eliminates star trails. \texttt{calib} astrometric plates are manually fit on the video data from both the Elginfield and Tavistock sites to ensure good quality of the astrometry and photometry.

Meteor position picks are done by manually defining the centroiding region position and radius (the semi-automated algorithm computes intensity-weighted centre of mass) and the photometry is done by masking which pixels belong to the meteor on each video frame. The astrometric picks are run through the meteor trajectory estimation code based on \cite{borovicka1990comparison} which uses a lines of sight method, and a heliocentric meteoroid orbit is computed. This initial solution is only used to decide whether the meteor warrants additional manual reduction.

\subsection{Manual reduction of narrow-field data}

\begin{figure}
  \includegraphics[width=\linewidth]{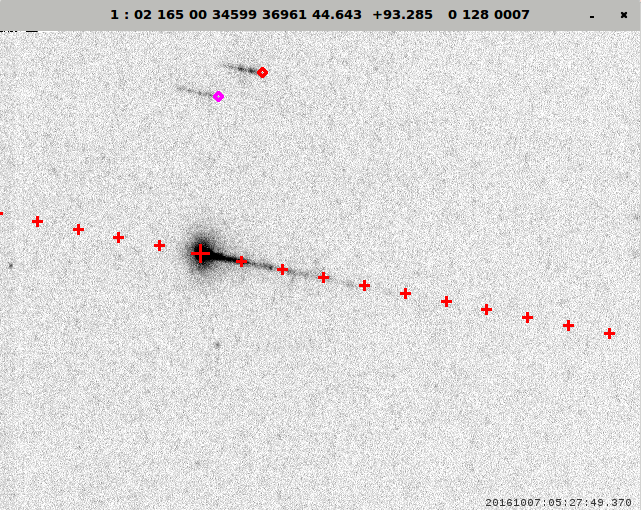}
  \caption{\texttt{mirfit} graphical user interface. A video frame with a meteor is shown, and the centroid pick for the current frame is marked with a large red plus sign, while astrometric picks on previous and subsequent video frames are marked with smaller red plus signs. Two diamonds mark the predicted positions of stars in the image, which become trailed as the mirrors track the meteor. The diamonds are the \texttt{exact} plate estimates of the two catalog star positions, and are at the end of each trail due to the camera timestamping each video frame at the end of the exposure.}
  \label{fig:mirfit_window}
\end{figure}

Narrow-field data is manually reduced using the \texttt{mirfit} software \citep[previously used in][]{subasinghe2017luminous}. With this software, the raw video frames and astrometric plates (\texttt{exact} and \texttt{scale}) which were created closest to the time of each event are loaded. The quality of the astrometric solution is confirmed by reverse mapping star catalog positions onto each video frame, and checking that they match the true positions of stars. Figure \ref{fig:mirfit_window} shows the \texttt{mirfit} graphical user interface and an example with two stars in the narrow field of view.

\begin{figure}
  \includegraphics[width=\linewidth]{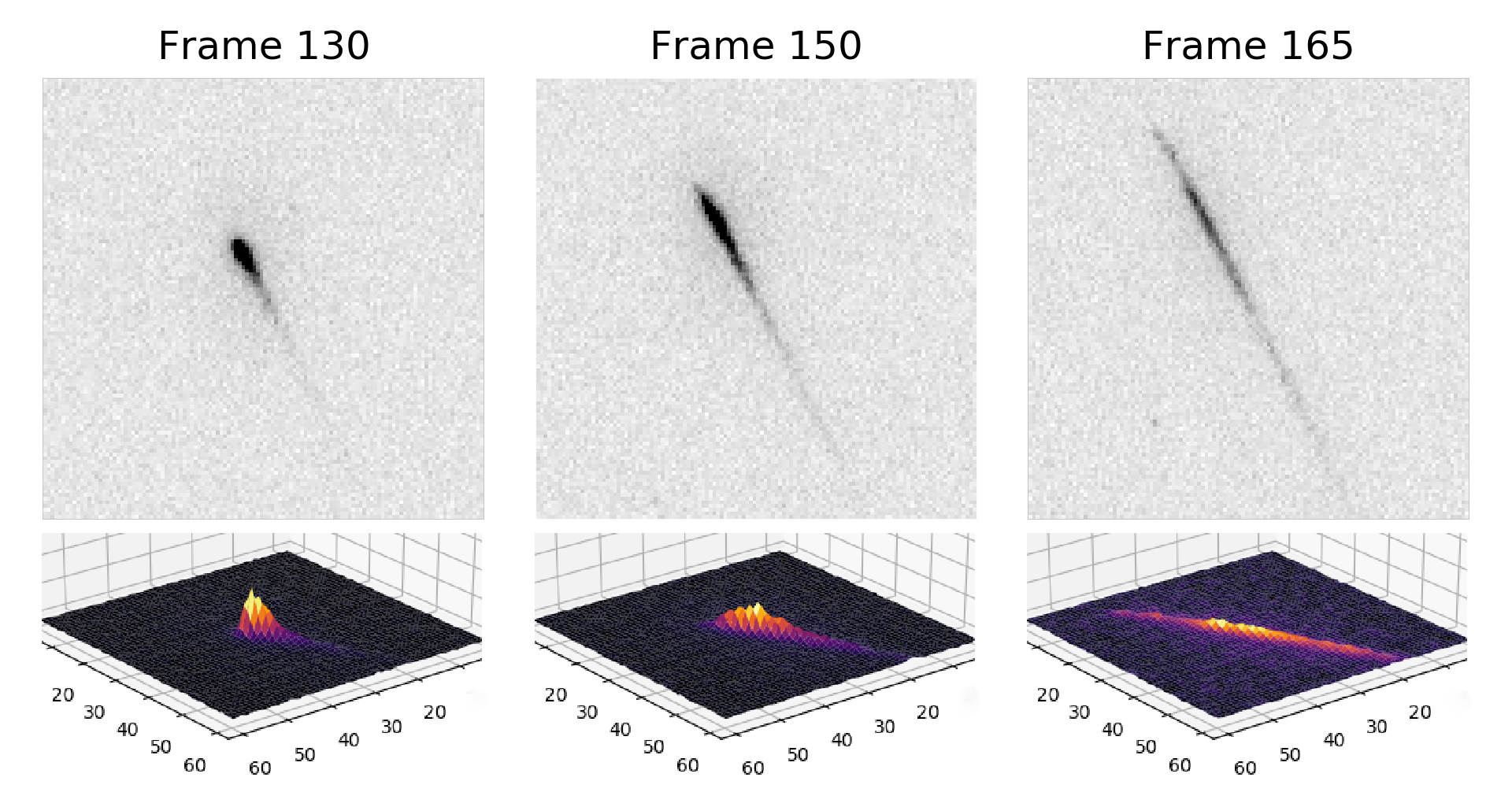}
  \caption{Three video frames from a meteor observed on May 17, 2017 at 04:54:26 UTC. The top plots show cropped video frames around the meteor, while the bottom plots shows their respective 3D intensity profiles.}
  \label{fig:img_vs_3d_profiles}
\end{figure}

Making meteor position picks on every frame is often difficult and subjective due to the complex morphology and fragmentation that may be present. In many cases, the precision of the meteor trajectory is limited by the morphology, regardless of the resolving power and precision of the CAMO tracking system. For example, Figure \ref{fig:img_vs_3d_profiles} shows a meteor that disintegrated into a long cylinder of luminous dust, making any consistent and precise astrometric picks after fragmentation impossible. However, we found that the best approach is to centroid on the most consistent leading fragment or feature that exists throughout the event as long as possible. This maximizes the number of picks, and better ensures a common feature is tracked from both sites. Sometimes the picks must be set manually to a pixel at the leading edge of the trail during fragmentation, as individual features cannot always be resolved.

\begin{figure}
  \includegraphics[width=\linewidth]{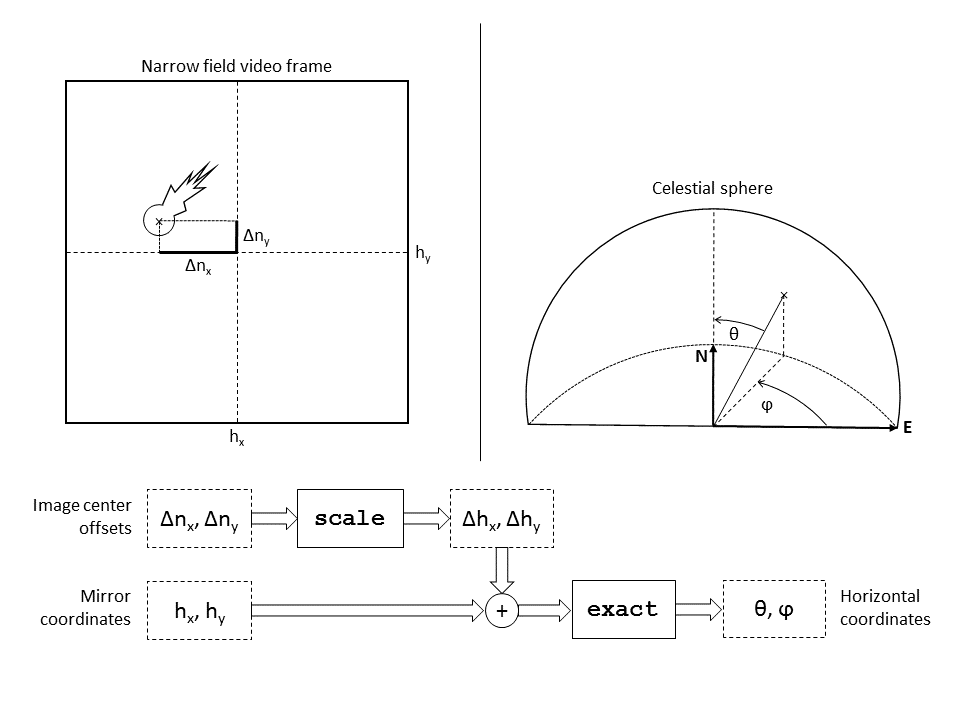}
  \caption{Schematic showing how the computation of narrow-field astrometry is performed.}
  \label{fig:plate_mapping_diagram}
\end{figure}

Computing celestial coordinates of observed meteors using narrow-field data is done in several steps. Figure \ref{fig:plate_mapping_diagram} shows a diagram of the procedure. First, assuming picks of meteor positions with coordinates $n_x, n_y$ were done on a particular narrow-field video frame, offsets $\Delta n_x, \Delta n_y$ from the narrow-field image center are computed. $\Delta n_x, \Delta n_y$ are then mapped into offsets in mirror units, $\Delta h_x, \Delta h_y$ using the \texttt{scale} plate. $\Delta h_x, \Delta h_y$ indicate the mirror encoder position offset required to centre the meteor in the narrow field. On average, a change of 3 mirror units will shift the narrow-field image by 1 pixel. 

As the narrow-field camera's frame rate is not phase synchronized with mirror position updates, the equivalent mirror pointing coordinates $h_x, h_y$ at the time the video frame was recorded are computed by linearly interpolating the recorded mirror positions in time. The offset in mirror units ($\Delta h_x$, $\Delta h_y$) is added to the actual mirror positions ($h_x$, $h_y$) at the frame time. Using the \texttt{exact} plate, the resulting mirror units are mapped onto the celestial sphere.  One pixel in the narrow-field image roughly corresponds to 6 arc seconds on the sky (\SI{3}{\metre} resolution at \SI{100}{\kilo \metre} range), and thus 1 mirror unit corresponds to about 2 arc seconds on the sky.

The narrow-field photometry is not used operationally because a meteor is usually spread over a large number of pixels in the image, which significantly reduces its signal to noise ratio, especially for fainter wakes. Also, there is a factor of 2 variation in sensitivity across the mirror field of regard, as the total mirror overlap is less on the edges of the field of regard. Nevertheless, photometry can be manually measured for individual events.

Once the narrow-field astrometric reduction is done from both sites, the meteor trajectory is computed using the Monte Carlo trajectory solver \citep{vida2019meteortheory}. This algorithm computes the radiant, heliocentric orbital elements, and associated uncertainties using the variance in the measured look angles as estimators for the precision of the measurements.

Due to the fact that meteoroids decelerate up to \SI{500}{\metre \per \second} prior to becoming detectable by CAMO \citep{vida2018modelling}, and that there exists a \SI{\sim 0.1}{\second} delay before the narrow-field tracking starts during which the meteoroid decelerates even more, we emphasize that all measurements of the initial velocity in this paper are surely underestimated. Thus, the stated uncertainties only reflect the measurement precision of the velocity at the start of narrow-field tracking, and not the absolute initial velocity accuracy at the top of the atmosphere. We do not apply the correction suggested by \cite{vida2018modelling} to avoid introducing potential biases and confusion in the measurements. As suggested by that work, in the future we will fit a meteoroid ablation and fragmentation model to our observations to invert the physical properties and the true initial velocity at the top of the atmosphere. Note that this approach will still potentially be limited by the suitability of the model, although CAMO can provide further constraints such as the meteoroid wake and the fragmentation details, in addition to the light curve and high-precision deceleration.

\section{Examples of reduced meteors} \label{sec:reduced_meteors}

In this section, we show three representative examples of ultra-high precision meteor trajectory solutions computed from CAMO narrow-field data and comment on their radiant and velocity accuracy. We show that the accuracy is mainly limited by meteor morphology; the three examples cover meteors having the most to the least favourable morphology.

\subsection{Morphologies allowing for high precision measurements}

\begin{figure}
  \includegraphics[width=\linewidth]{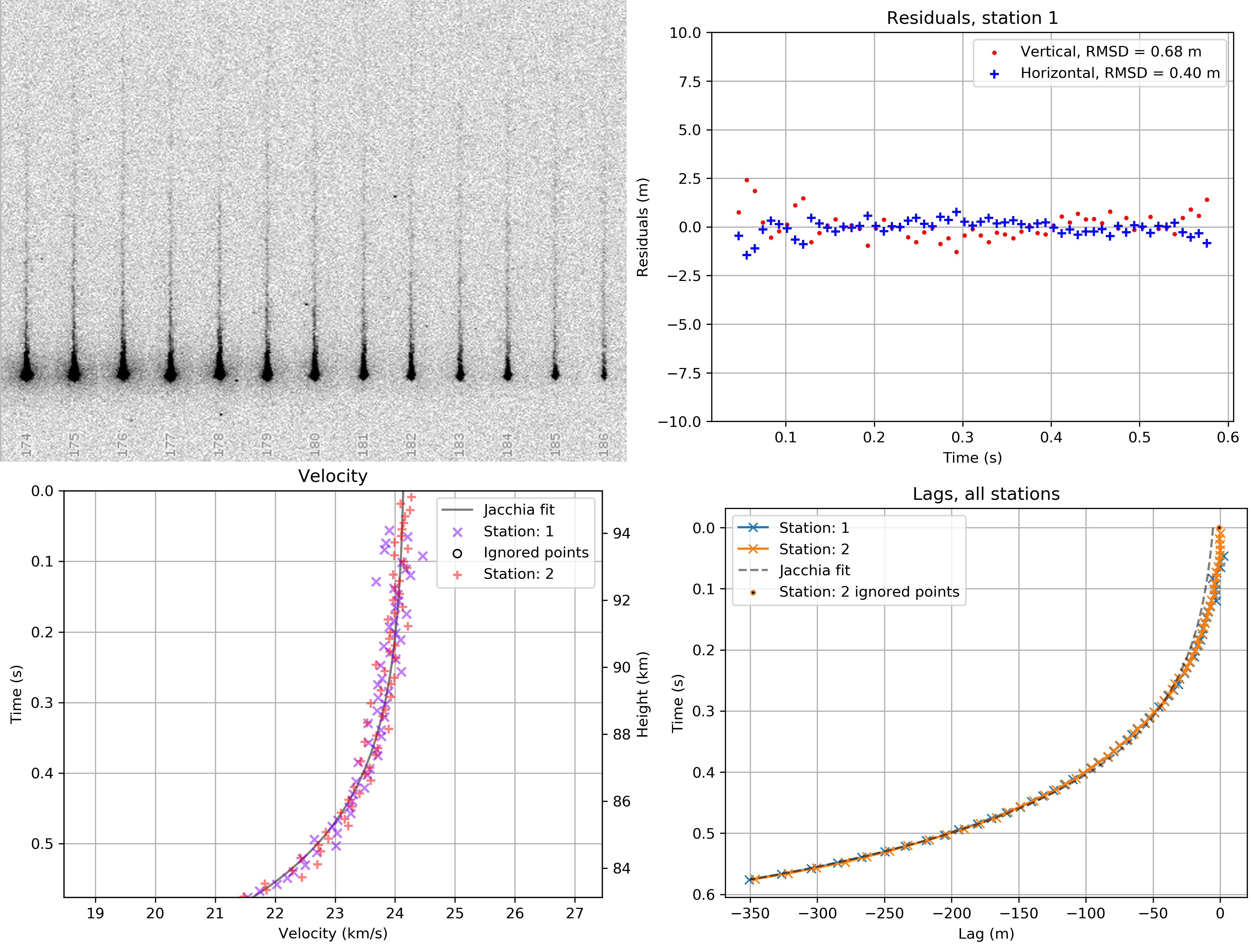}
  \caption{Reduction of a meteor observed on October 7, 2016 at 05:27:49 UTC. Upper left: A composite image of the last 13 narrow-field video frames rotated and cropped so that the leading edge is aligned in every frame. The time progresses from left to right at \SI{10}{\milli \second} increments. Upper right: Spatial fit residuals from the Tavistock site. Lower left: Point-to-point velocities. Lower right: Lag of the leading edge compared to a fixed velocity model. The ``Jacchia fit" shows a fitted exponential deceleration model given in equation \ref{eq:jacchia_lag_func}. }
  \label{fig:20161007_052749_camo_event}
\end{figure}

Figure \ref{fig:20161007_052749_camo_event} shows a composite image of a sporadic meteor observed by the CAMO tracking system on October 7, 2016, with the meteor being well tracked from both sites. As seen in the figure, the spatial fit residuals are below one meter (the corresponding angular residuals are $\sim 1$ arc second), and the point-to-point velocities are very compact and show obvious smooth deceleration. The lag, defined as the ``distance that the meteoroid falls behind an object with a constant velocity that is equal to the initial meteoroid velocity'' \citep{subasinghe2017luminous}, matches well between both stations, an indication of a good trajectory solution \citep{vida2019meteortheory}. The meteor showed only continuous fragmentation and no gross fragmentation; this favourable morphology contributed to the quality of the trajectory solution.

In Table \ref{tab:example_meteor_orbits} we give the radiant and osculating orbital elements computed using the Monte Carlo trajectory solver \citep{vida2019meteortheory}. The stated uncertainties are small, but there are several caveats. First, the compensation for deceleration prior to detection was not done, thus the initial velocity may be underestimated as much as \SI{500}{\metre \per \second} \citep{vida2018modelling}. Consequently, the stated velocity measurement uncertainty gives the precision, not the accuracy. Second, \cite{vida2019meteorresults} have shown that radiant uncertainties for CAMO are usually underestimated by a factor of 3 to 4 with this solver based on comparison with simulations. Third, due to the time needed for the narrow-field tracking to begin, the initial velocity is even further underestimated. Thus the real radiant accuracy is probably $\sim \ang{0.025}$, well within the minimum precision of \ang{0.1} necessary to measure the true physical dispersions of meteor showers \citep{vida2019meteorresults}. As for the initial velocity, to reconstruct the original value without the deceleration from various sources, an meteoroid ablation and fragmentation model will have to fit to observations \citep{vida2018modelling}.

\subsection{Morphologies with deteriorating measurement precision}

\begin{figure}
  \includegraphics[width=\linewidth]{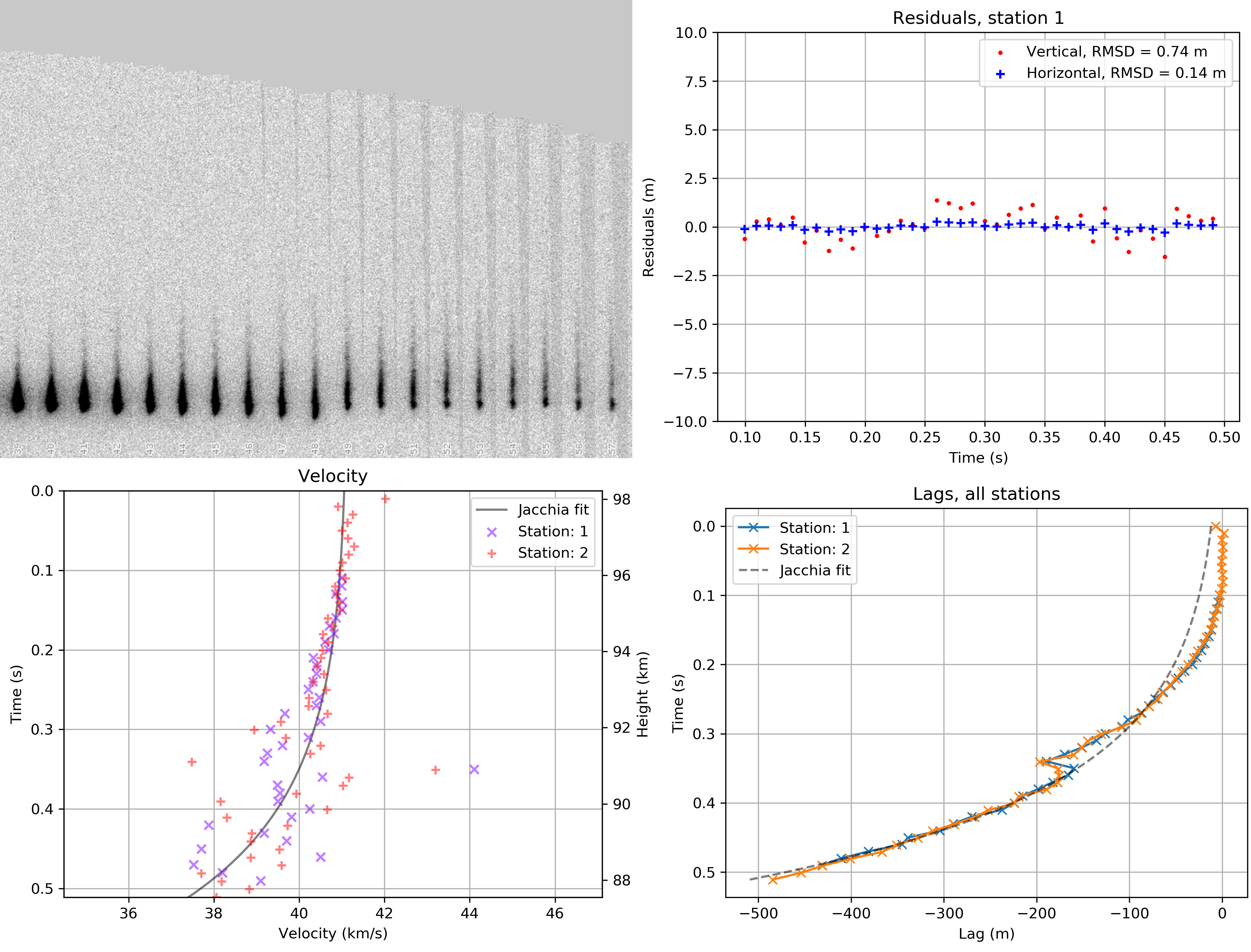}
  \caption{Reduction of a meteor observed on August 10, 2019 at 06:19:57 UTC. Upper left: A composite image of the middle 19 narrow-field video frames rotated and cropped so that the leading edge is aligned in every frame. The morphology change occurs halfway though the shown frames. The time progresses from left to right. Upper right: Spatial fit residuals from the Tavistock site. Lower left: Point-to-point velocities. Lower right: Lag with an inflection corresponding to the change in morphology which is visible from both stations. }
  \label{fig:20190810_061957_camo_event}
\end{figure}

Figure \ref{fig:20190810_061957_camo_event} shows a Southern $\delta$ Aquariid meteor observed on August 10, 2019 at 06:09:57 UTC. This meteor had a sudden change in morphology halfway through the observation caused by an increase in the rate of continuous ablation. Prior to the change, the meteor had compact morphology with a short wake. The whole meteor was centroided prior to the morphology change during the manual reduction, excluding the short wake, which produced robust astrometry picks and well-matched velocities from both sites. After the morphology change, the wake became longer and the meteoroid morphology became elongated, showing a leading fragment at the front. At that point the leading fragment was followed, but due to the lower signal to noise ratio and interference from released grains and the wake, the picks were less consistent, which caused a large spread in point-to-point velocities. The change can be seen in the lag as a sudden shift back towards the zero lag line, as the reference point moved forward to the leading fragment by a fixed amount. However, this change did not influence the trajectory fit residuals as the leading fragment did not have a transverse velocity component, though this is not always the case \citep[e.g.][]{stokan2014transverse}. Table \ref{tab:example_meteor_orbits} provides the orbital elements and uncertainties for this event. The uncertainties in the radiant and the orbital elements are larger than for the previous event, mainly due to the larger pick scatter in the second half of the meteor trajectory. In contrast, the geocentric velocity uncertainty remained low because the initial velocity used for orbital computation is found using data from the first part of the meteor, in this case prior to the morphology change.

\subsection{Morphologies which severely limit measurement precision}

\begin{figure}
  \includegraphics[width=\linewidth]{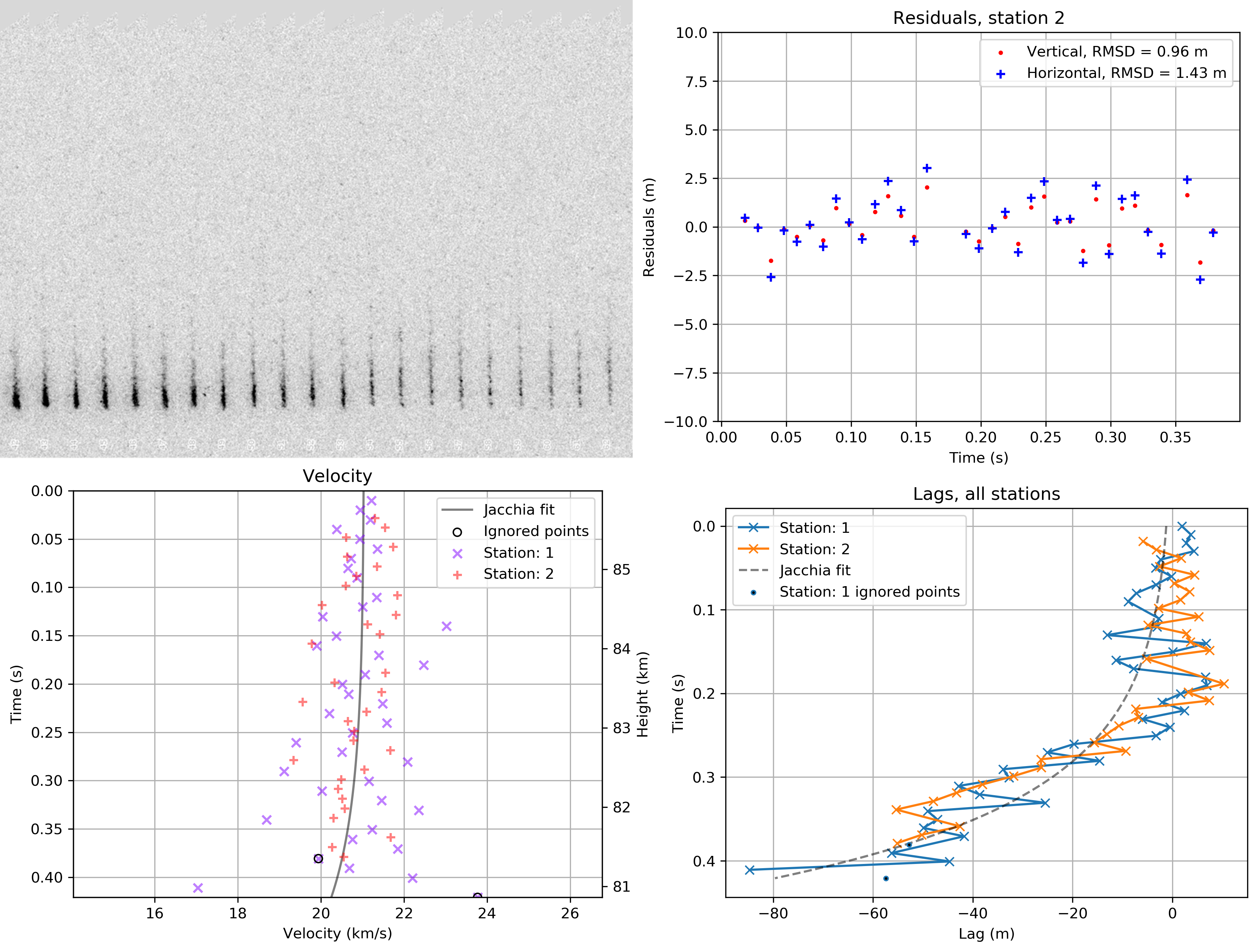}
  \caption{Reduction of a meteor observed on August 31, 2019 at 09:12:42 UTC. Upper left: A composite image of 21 selected narrow-field video frames rotated and cropped so that the leading edge is aligned in every frame. The time progresses from left to right. Upper right: Spatial fit residuals from the Tavistock site. Lower left: Point-to-point velocities. Lower right: Lag in distance relative to a constant velocity. }
  \label{fig:20190831_091242_camo_event}
\end{figure}

As a final end-member example, Figure \ref{fig:20190831_091242_camo_event} shows a meteor on an asteroidal orbit with a probable higher bulk density than the earlier cases, judging from the height range and small deceleration. It exhibited complex morphology (no leading fragment, extended meteor luminosity mostly consisting of a wake), and as a consequence, it was difficult to make consistent position picks. This is reflected in the higher scatter of spatial residuals, velocity, and lag. In this particular case, the meteor morphology was the limiting factor in achieving better astrometric precision. The radiant and the orbital elements are given in Table \ref{tab:example_meteor_orbits}. All uncertainties are larger than for the two previous events due to the larger scatter in astrometric picks. Assuming that the radiant uncertainty was underestimated by a factor of 4, the true total radiant uncertainty is $\ang{\sim 0.2}$, which may not be sufficiently accurate to measure true physical radiant dispersions of compact meteor showers.

\begin{table*}[t] %floating table at the top of the page
	\caption{The geocentric radiant in J2000 and corresponding heliocentric orbital elements of example meteor events with different morphologies. The radiant and velocity precision worsens for meteoroids with more fragmentation. The stated uncertainties are one sigma errors and only state the fit precision, not the absolute accuracy. The initial velocity is underestimated due to deceleration prior to detection, and due to the time needed for the mirror to lock on to the meteor. }
	\label{tab:example_meteor_orbits} % is used to refer this table in the text
	\centering % used for centering table
	
	\begin{tabular}{l | S[table-format=3.5] l | S[table-format=3.5] l | S[table-format=3.5] l}
	\hline\hline % inserts double horizontal lines
	
	& \multicolumn{2}{c|}{October 7, 2016 meteor} & \multicolumn{2}{c|}{August 10, 2019 meteor} & \multicolumn{2}{c|}{August 31, 2019 meteor} \\ % table heading

	\hline % inserts single horizontal line
    
    Description & \multicolumn{2}{c|}{Short wake} & \multicolumn{2}{c|}{Fragmentation half-way} & \multicolumn{2}{c}{Long wake, low SNR} \\ % table heading	
	$\alpha_g$ (deg) &  11.4761  & $\pm 0.0015 $  & 348.6155  & $\pm 0.0033 $   & 337.2336  & $\pm 0.0086$ \\
	$\delta_g$ (deg) & +15.9913  & $\pm 0.0057 $  & -14.0499  & $\pm 0.0068 $   & +12.021   & $\pm 0.045$ \\
	$v_g$ (km/s)     &  21.474   & $\pm 0.0010 $  &  39.48892 & $\pm 0.0008 $   &  17.805   & $\pm 0.018$ \\
	a (AU)           &  1.9154   & $\pm 0.0001 $  &  2.55591 & $\pm 0.0005 $    &  1.33822  & $\pm 0.0005$ \\
	q (AU)           &  0.556308 & $\pm < 0.0001$ &  0.100421 & $\pm < 0.0001$  &  0.59353  & $\pm 0.0003$ \\
	e                &  0.709573 & $\pm < 0.0001$ &  0.960710 & $\pm < 0.0001$  &  0.55648  & $\pm 0.0004$ \\
	i (deg)          &  7.5026   & $\pm 0.0044 $  & 26.368    & $\pm 0.0204 $   & 12.390    & $\pm 0.037$ \\
	$\omega$ (deg)   & 273.9063  & $\pm 0.0032 $  & 147.0144  & $\pm 0.0070 $   & 278.775   & $\pm 0.029$ \\
	$\Omega$ (deg)   & 194.15640 & $\pm < 0.0001$ & 317.09080 & $\pm < 0.0001$  & 157.43669 & $\pm < 0.0001$ \\
	
	\hline %inserts single line
	\end{tabular}
\end{table*}

\section{Meteoroid compressive strengths derived from direct observations of gross fragmentation} \label{sec:strengths_measurements}

In contrast to optical meteor systems used for previous estimates of meteoroid compressive strengths (see the summary in Section \ref{subsec:strengths_atm}), CAMO can directly observe gross fragmentation of meteoroids. Earlier we described the CAMO data calibration and reduction tools, and demonstrated how meteor morphology limits the ultimate trajectory accuracy. In this section, we discuss those CAMO meteors which show gross fragmentation of the main meteoroid body into several discrete fragments, a process which we assume results from structural failure of the meteoroid under the action of atmospheric dynamic pressure. Meteors with this morphology make up about 5\% of all observed meteors with CAMO, and this morphology is not correlated with orbital type \citep{subasinghe2016physical}.

These events often have the least accurately defined astrometry because there is no single consistent point of reference that can be tracked. Frequently at the beginning of the luminous track the meteor may resemble a single object, but the amount of continuous fragmentation is usually high and any further consistent astrometric picks become impossible once gross fragmentation occurs. 
Thus, we only use the wide-field data to compute the reference trajectory, and we project the narrow-field astrometric picks of individual fragments onto it to determine their dynamics. High-precision measurement of fragment deceleration allows us to compute precise values of the aerodynamic ram pressure as we know the height and speed at each frame. By observing when fragmentation occurs in the narrow field imagery, we can estimate precise values of meteoroid compressive strength.

\subsection{Sensitivity analysis} \label{subsec:strengths_sensitivity_analysis}

First we examine the uncertainty of individual parameters used to compute the dynamic pressure from equation \ref{eq:dyn_press}. For the drag coefficient, we use $\Gamma = 1$, a value appropriate for a sphere, which is a common assumption for meteoroid ablation in free molecular flow \citep{fisher2000meteoroids, campbell2004model, borovicka2007atmospheric, vida2018modelling}. In fact, many works even exclude the drag coefficient from the equation, implicitly assuming it is unity \citep[e.g.][]{trigo2006strength, blum2014comets}.

Observations of the morphology of cometary dust particles by \cite{hilchenbach2016comet} indicate that meteoroid components are oblate spheroids, although we are not aware of any works showing detailed shape analysis. If the axial ratios of spheroid meteoroid components were to vary from 0.5 to 1.0, drag coefficients may also vary within a factor of two \citep{list1973aerodynamics}. If that was the case, and the meteoroids were not rotating, we would expect to see a comparable variation in dynamic pressures at points of fragmentation among fragments of one meteoroid, assuming a fixed velocity, atmospheric mass density, and homogeneous strength. 

In section \ref{subsec:july21_event_dyn_press} we further discuss the possibility of drag coefficient variation using direct observations of fragments of one meteoroid. $\Gamma$ also varies with the Reynolds number, but \cite{thomas1951physical} show that spherical meteoroids moving in a highly rarefied gas and at hypersonic speeds have $\Gamma \sim 1$, thus in this work we fix it to unity.

For our events, uncertainty in $P_{dyn}$ is not driven by the uncertainty in the velocity measurement. \cite{vida2019meteorresults} have shown that initial velocities can be reliably measured to within \SI{0.5}{\kilo \metre \per \second}. Even if one assumes a low initial velocity of only \SI{10}{\kilo \metre \per \second}, the maximum error in dynamic pressure is only 10\%.

The last term required to compute the dynamic pressure is atmospheric mass density. The most sophisticated atmosphere mass density model available to date is the NRLMSISE-00 model \citep{picone2002nrlmsise} which gives the atmosphere mass density as a function of geographical location, time, and solar activity. The time component takes the influence of seasonal changes into account. The solar activity is modelled by using the \SI{10.7}{\centi \metre} solar flux $F_{10.7}$ which slowly changes with the 11 year solar cycle, although it can change dramatically on shorter time scales due to the evolution of active regions and solar flares \citep{tapping201310}. \cite{picone2002nrlmsise} show that the influence of changes in $F_{10.7}$ can cause the air mass density in the upper atmosphere (\SI{> 600}{\kilo \metre} altitude) to change with an amplitude of half an order of magnitude.

Comprehensive models like The Whole Atmosphere Community Climate Model (WACCM) \citep{Qian:2013_WACCM_mass_density} or the Spectral Mesosphere/Lower Thermosphere Model (SMLTM) \citep{AKMAEV20061879_mass_density_trend} provide some insight into the mass density changes over time due to greenhouse gas cooling in the mesosphere or solar effects. Recently, these trend studies were updated using WACCM-X \citep{Solomon:2019_WACCM-X_density_trends}, echoing the previously reported density changes at the Mesosphere and Lower Thermosphere (MLT). It was found that changes in mass density in the MLT are larger as compared to the altitudes below and directly above the MLT (\SIrange{150}{200}{\kilo \metre}).

From meteor radar observations with the co-located CMOR (Canadian Meteor Orbit Radar) a neutral air density change of approx. 6\% per decade was obtained \citep{Stober:2014_neutral_density_variation}, which corresponds well to the WACCM and SMLTM results. Additionally, a solar cycle duration response of 2-3\% in the neutral air density was estimated. Other meteor radar studies using the meteor ablation altitude \citep{Jacobi:2011_meteor_alt_sol_cycle, LIMA:2015139_meteor_heights_sol_cycle, Liu:2017_meteor_height_variations_Mohe} or from vertical profiles of the ambipolar diffusion measurements \citep{Yi:2019_density_variability_ambipolar_diff} have also estimated the seasonal variability of the neutral air density. 

However, it is the short term variability of the neutral air density induced by atmospheric waves that is most germane for uncertainty analysis for compressive strength estimation from CAMO. \cite{Stober:2012_neutral_air_density_PW} explored the magnitude of the neutral air density at heights of interest for CAMO and showed that they can vary due to planetary waves during the winter season 2009/2010 using three meteor radars across Europe.

We investigated the short term variability of the neutral air density using meteorological fields from the NAVGEM-HA (Navy Global Environmental Model- High Altitude) numerical weather prediction system \citep{hogan2014navy,McCormack:2017,Eckermann:2018}. NAVGEM-HA combines a global forecast model of the atmosphere with a 4DVAR hybrid data assimilation scheme \citep{kuhl:2013} to produce global atmospheric specifications for a given time period extending from the surface to \SI{\sim 116}{\kilo \metre} altitude. NAVGEM-HA assimilates a over 3 million ground-based and satellite-based observations every 6 hours. In the altitude region from \SIrange{20}{100}{\kilo \metre} the primary observation sources are temperature, ozone, and water vapor retrievals from the Microwave Limb Sounder (MLS) on board the Aura satellite and temperature retrievals from the SABER instrument on board TIMED. The NAVGEM-HA output used in this study consists of global 6-hourly wind, temperature, and geopotential height fields on a \ang{1} latitude/longitude grid over 74 vertical levels from 1 January to 31 December 2010. At the upper two model levels (above \SI{\sim 95}{\kilo \metre} altitude), enhanced horizontal diffusion (i.e., a ``sponge layer'') is applied to reduce wave reflection \citep{McCormack:2015_sponge_layer_NAVGEM}. A validation of the NAVGEM-HA fields at the CAMO Tavistock site can be found in \cite{Stober:2019_NAVGEM-HA}.

\begin{figure}
  \includegraphics[width=\linewidth]{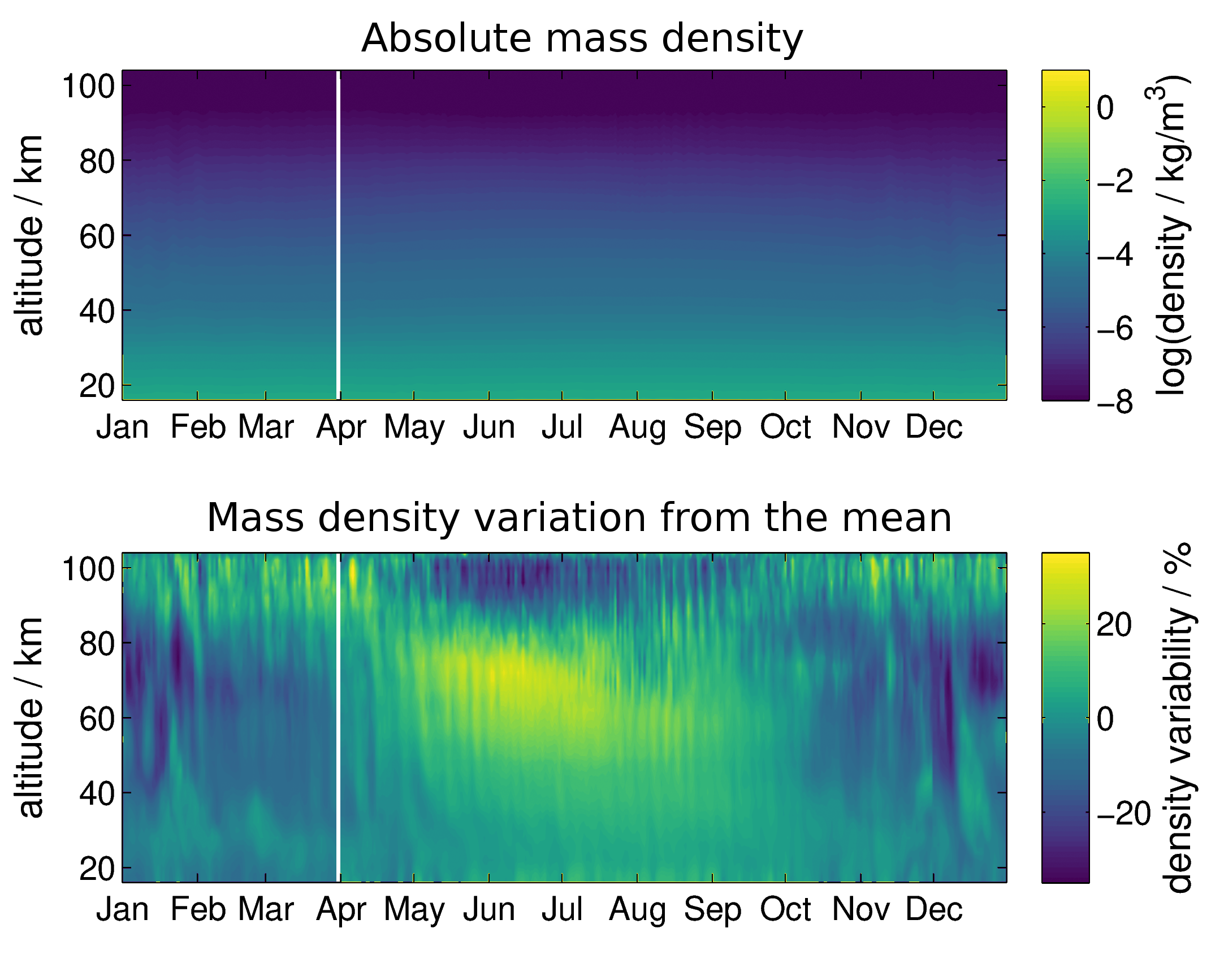}
  \caption{Measurement of atmosphere mass density using NAVGEM-HA above the CAMO Tavistock site. The upper panel shows absolute values of the mass density as daily mean values for the year 2010. The lower panel visualizes the relative density variability variation from the mean.}
  \label{fig:navgem_atm_densities}
\end{figure}

The annual mass density variation at the Tavistock site based on NAVGEM-HA is shown in Figure \ref{fig:navgem_atm_densities}. The upper panel shows the absolute density values as contour plot with logarithmic scaling, and the lower panel shows the relative variability in percent. We computed a median density value for each geopotential altitude for the whole year of 2010 and used it to normalize all values. Hence, the variability plot not only contains the seasonal variability but also the atmospheric waves. From the plots, it is apparent that the mass density at meteor heights can vary by up to $\pm 25\%$ on short time scales. It is therefore the main driver of the uncertainty in the dynamic pressure. We adopt this $\pm 25\%$ value as representative of the air mass density uncertainty and use it in what follows to estimate all dynamic pressure measurement uncertainties.

\subsection{July 21, 2017 event} \label{subsec:july21_event}

\begin{figure}
  \includegraphics[width=\linewidth]{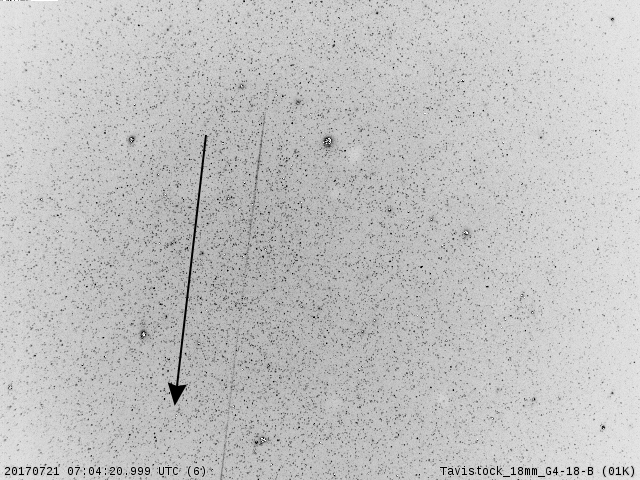} \hfill
  \includegraphics[width=\linewidth]{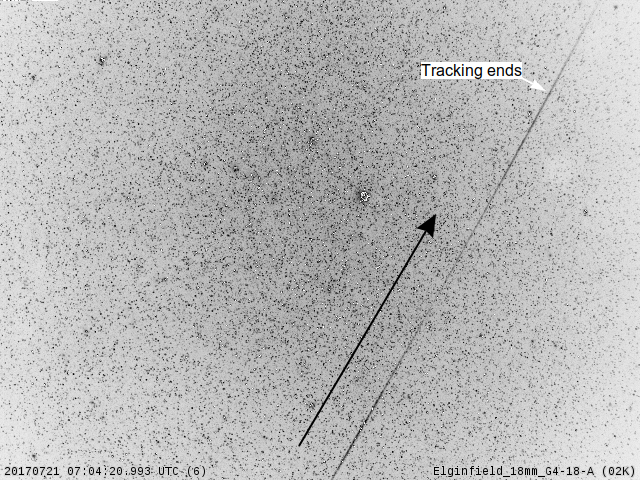} 
  \caption{Grey-inverted co-added video frames of the July 21, 2017 event. The images are heavily speckled by intensifier shot noise due to co-adding; individual video frames have a much better signal to noise ratio. Top: Tavistock wide-field camera. Bottom: Elginfield wide-field camera. The black arrow indicates the direction of flight, and the white arrow indicates the point on the trajectory where the tracking from the Tavistock site ended.}
  \label{fig:july21_widefield}
\end{figure}

We begin with a specific case study of an unusual fragmenting meteoroid observed on July 21, 2017. It  had a very shallow entry angle of \ang{8} degrees and was observed for almost 4 seconds and shows a clear double-peaked lightcurve. It was observed by the wide-field cameras from both sites almost in its entirety, but was only well tracked by the Tavistock narrow-field camera - the tracking parameters were not well estimated from Elginfield (possibly due to the long wake) where it was only tracked for a few frames before it exited the field of view. Figure \ref{fig:july21_widefield} shows the co-added video frames captured by the wide-field cameras from both sites.

\subsubsection{Morphology} \label{subsec:july21_even_morphology}

\begin{figure*}
    \centering
    \includegraphics[width=\textwidth,height=\textheight,keepaspectratio]{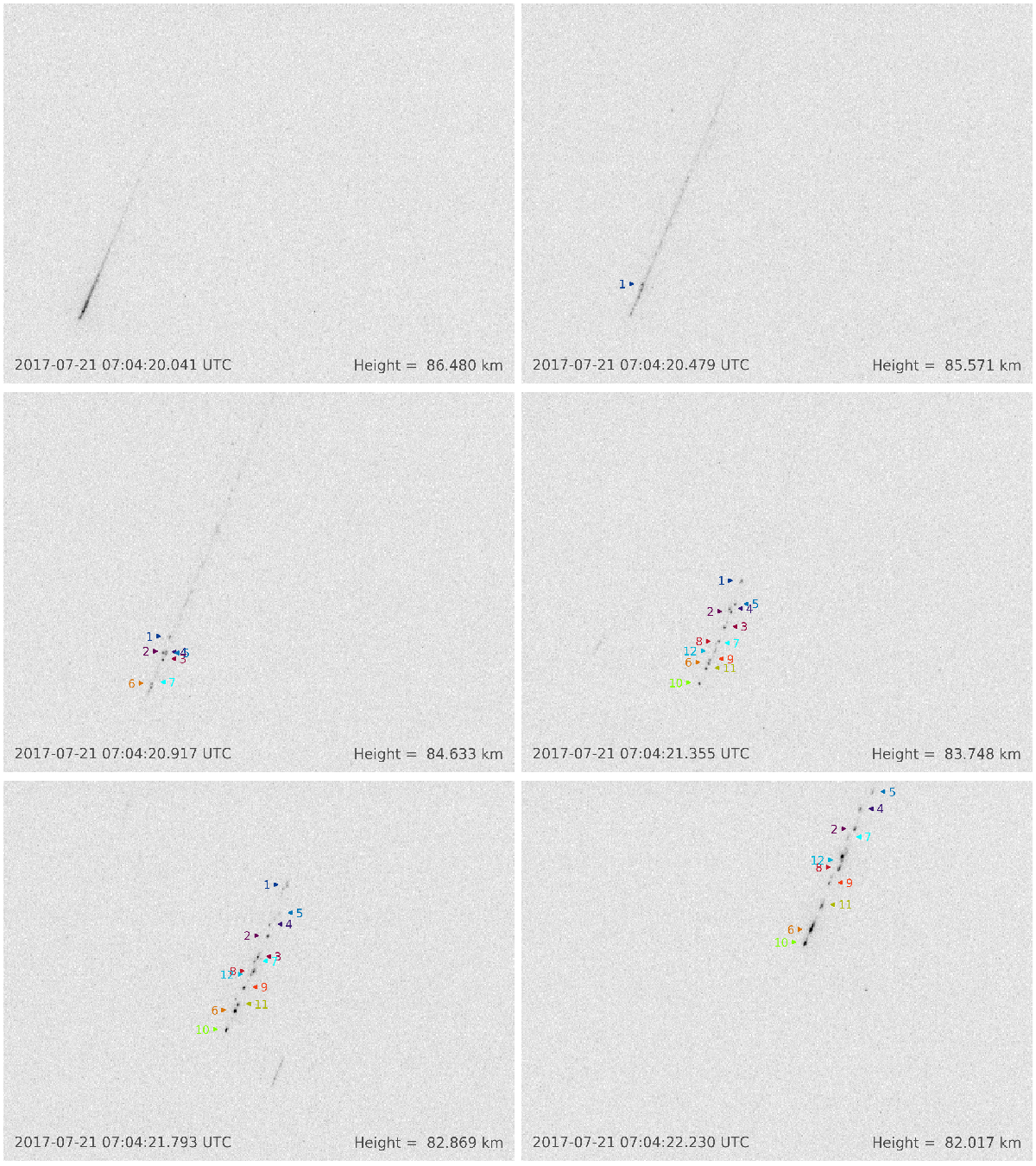} 
    \caption{Mosaic of six narrow-field video frames from the Tavistock site for the July 21, 2017 meteor. Each discrete fragment has been tracked and labeled with a unique number.}
    \label{fig:july21_mosaic}
\end{figure*}

The tracking at Tavistock started 0.45 seconds after it was initially observed in the wide-field camera, as the meteor was below the automated detection threshold before that time. Figure \ref{fig:july21_mosaic} is a mosaic of six narrow-field video frames from Tavistock which shows the morphological evolution. When the tracking started, an extended wake could be seen in the wide-field video. For the complete narrow-field video, see the supplementary materials or \url{http://meteor.uwo.ca/~dvida/IMC2017/20170721_tavis_narrow.gif}.

\begin{figure}
  \includegraphics[width=\linewidth]{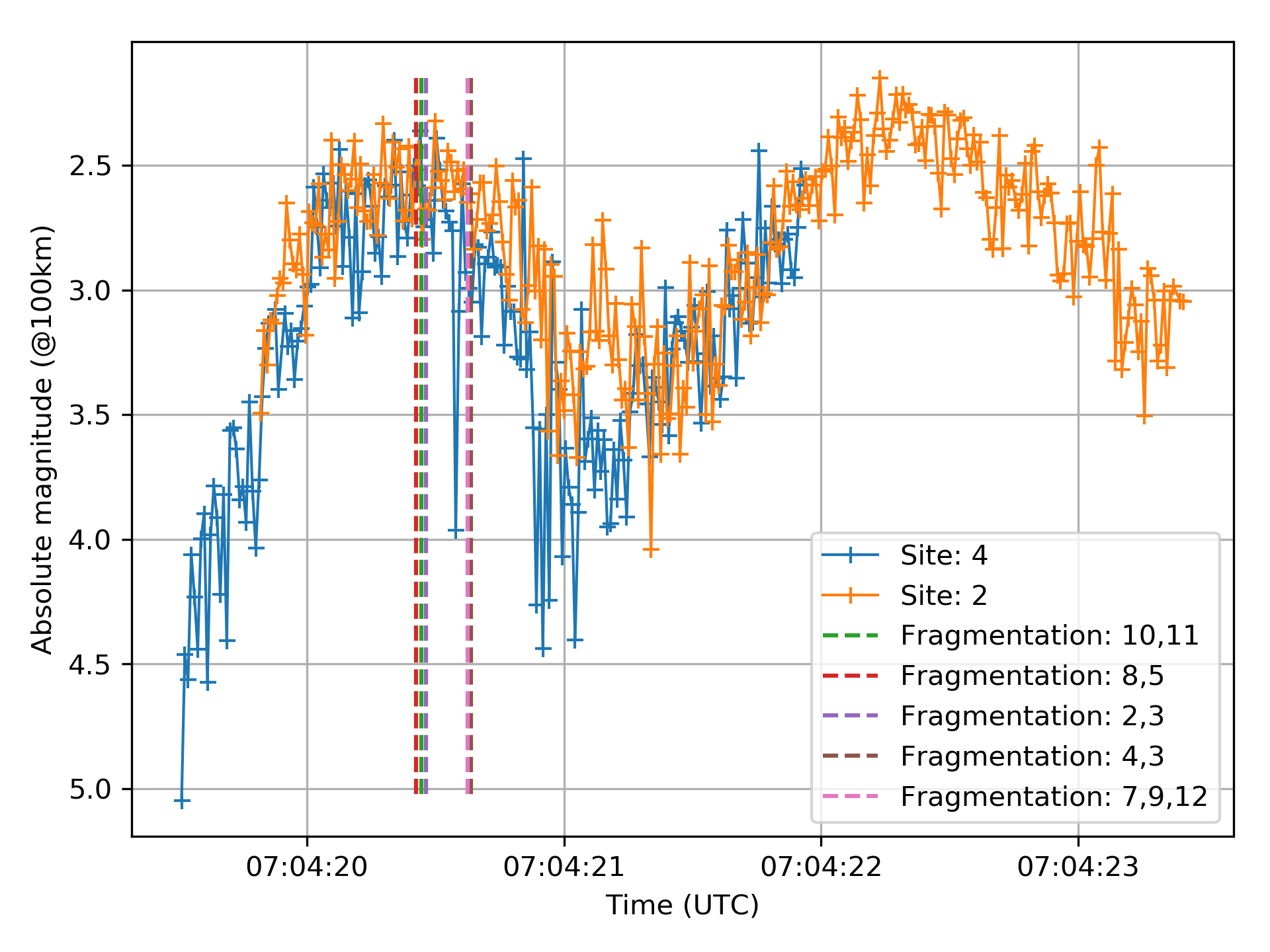} 
  \caption{Wide-field light curve of the July 21 fragmenting event. Station 2 is Elginfield, station 4 is Tavistock. Vertical lines show the moments of gross fragmentation visible in the narrow field imagery.}
  \label{fig:july21_lightcurve}
\end{figure}

The narrow-field video showed that the wake consisted of unresolvable grains (or dust) lagging behind several fragments - the fragments themselves were also enveloped in the dust. Starting 1.4 seconds after the tracking began, the dust was completely gone, leaving 12 discrete fragments visible. During this time, the fragments noticeably decelerated and some showed transverse motion. Several fragments with lower deceleration, which were always brighter and presumably more massive, overtook fainter fragments. During this period devoid of wake, the wide-field video shows a significant dip in the brightness of the meteor as a whole, as shown in Figure \ref{fig:july21_lightcurve}.

\begin{figure}
  \includegraphics[width=\linewidth]{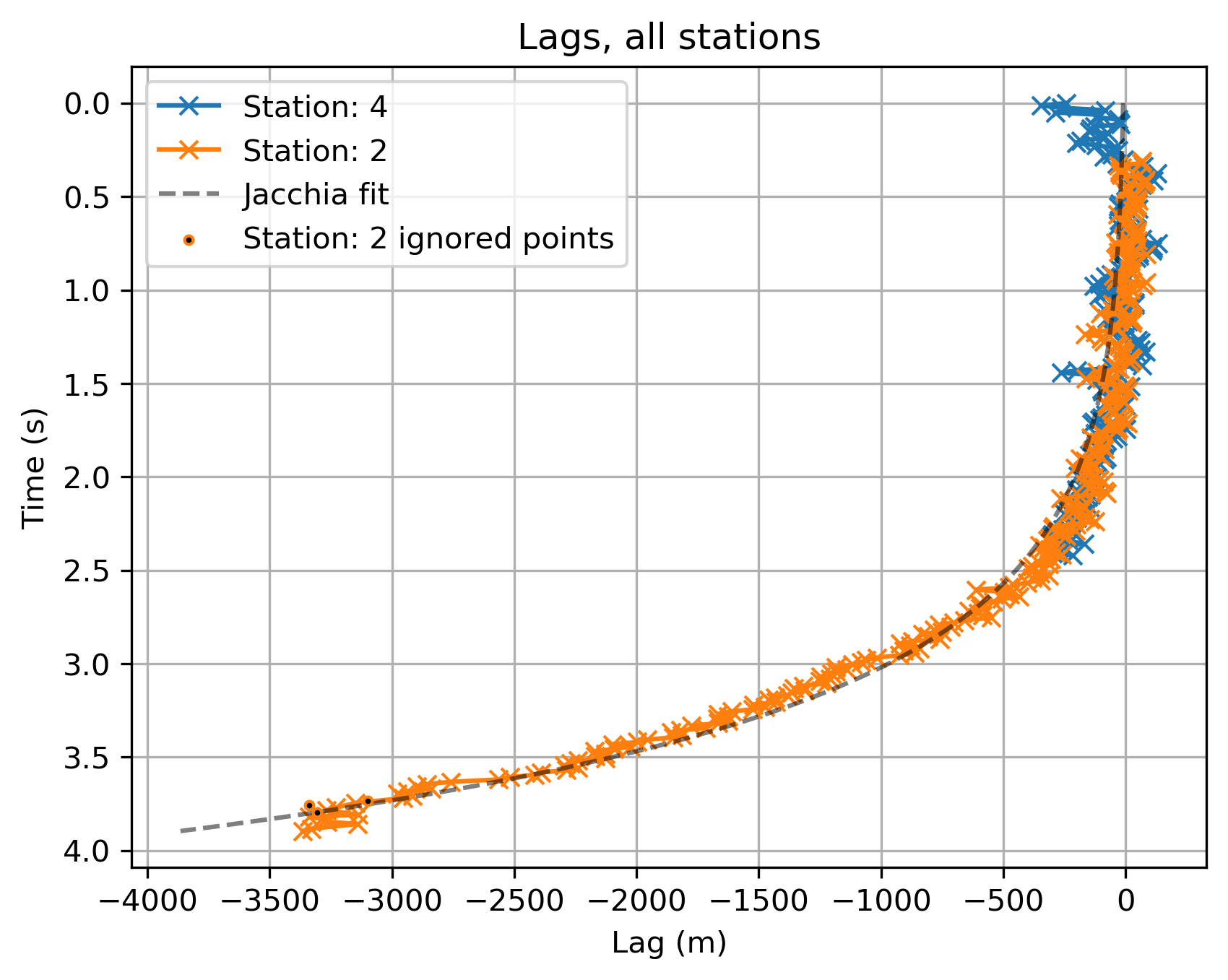} 
  \caption{Wide-field lag for the leading edge of the visible meteor. Station 2 is Elginfield, station 4 is Tavistock.}
  \label{fig:july21_lag}
\end{figure}

After one more second, the fragments developed short wakes and themselves disintegrated, and the total brightness increased again. At this point, the measurements from the wide-field camera show that the bulk of the meteoroid started to rapidly decelerate, as shown in Figure \ref{fig:july21_lag} (around 2.5 seconds), and that this disintegration produced a second peak in the light curve. Meteors with double peaked light curves were investigated by \cite{roberts2014meteoroid} and \cite{subasinghe2019properties}. However,  the observed behaviour of this event does not match any of their proposed meteoroid ablation or fragmentation mechanisms.

\subsubsection{Radiant and orbit}

Because narrow-field data was only available from the Tavistock site, we used wide-field data from both sites for trajectory and orbit calculation. Due to its long duration, the meteor experienced significant bending of the trajectory from a straight line due to gravity. This is taken into account using the \cite{vida2019meteortheory} meteor trajectory estimation method which we use here. 

\begin{figure}
  \includegraphics[width=\linewidth]{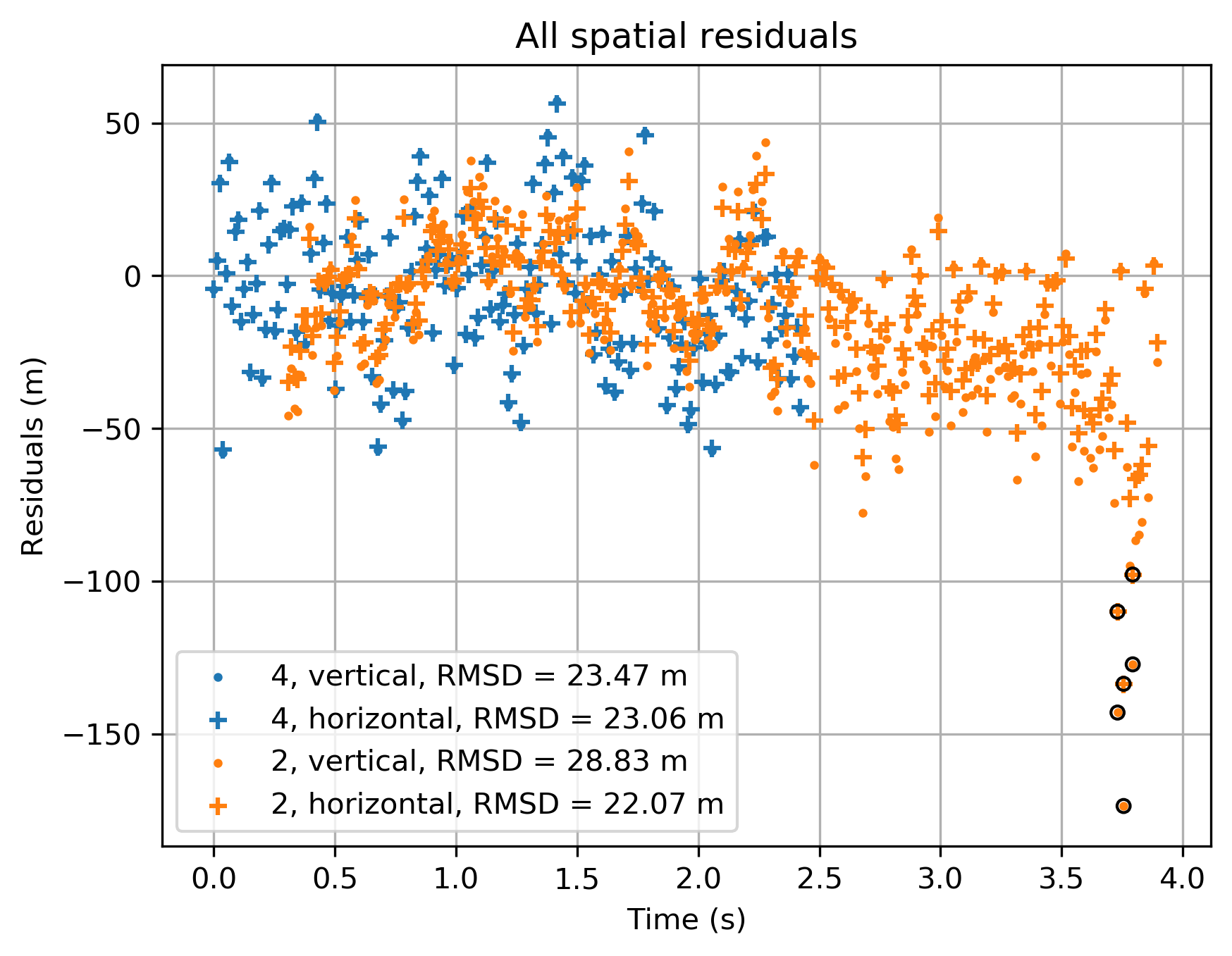} 
  \caption{Wide-field camera trajectory fit residuals. The bending of the trajectory due to gravity is visible. This should be $\sim \SI{50}{\metre}$ for 3.5 seconds of flight. Note that the residuals are computed with respect to a straight line aligned with the radiant, and not a curved trajectory; thus the vertical residuals show an offset near the end. Black circles indicate $3 \sigma$ outliers which were excluded from the trajectory fit.}
  \label{fig:july21_fit_residuals}
\end{figure}

Figure \ref{fig:july21_fit_residuals} shows the total trajectory fit residuals computed with respect to the radiant line, showing how significant the deviation from the straight line approximation is in this case. The average trajectory fit residuals from both sites were around \SI{25}{\metre}, which translates to about 1 arc minute.

From the wide field imagery light curve, we computed a photometric mass of $m = \SI{0.27}{\gram}$ using the bolometric power of a zero-magnitude meteor $P_{0m} = \SI{840}{\watt}$ \citep{weryk2013simultaneous} and a dimensionless luminous efficiency $\tau = 0.7\%$ \citep{campbell2013high} in the red bandpass. Because the end of the meteor was not observed, this mass is a lower limit. Furthermore, due to the uncertainty of the luminous efficiency, the mass uncertainty is at least a factor of 2 \citep{subasinghe2018luminous}. Assuming a bulk density of \SI{700}{\kilogram \per \cubic \metre}, the meteoroid had a diameter of \SI{9 \pm 2}{\milli \metre}.

The meteoroid entered the atmosphere at an angle from the horizontal of only \ang{7.8}. It was first observed at a height of \SI{87.849}{\kilo \metre} and it exited the wide camera field of view \SI{3.89}{\second} later at a height of \SI{79.882}{\kilo \metre}. The velocity at the beginning was $15.9626 \pm \SI{0.0051}{\kilo \metre \per \second}$, although it had certainly decelerated from its true pre-atmosphere velocity due to the low entry angle, low velocity, and small size \citep{vida2018modelling}. 

To quantify the amount of deceleration prior to detection, we used the single-body version of the ablation model of \cite{campbell2004model} to simulate the meteoroid and compute the deceleration from the top of the atmosphere (assumed at \SI{180}{\kilo \metre}) until it was detected by the wide-field cameras. The simulation roughly reproduces the observed conditions at the point of detection assuming a mass $m = \SI{0.17}{\gram}$, bulk density $\rho = \SI{700}{\kilogram \per \cubic \metre}$, heat of ablation $Q = \SI{4600}{\kilo \joule \per \kilogram}$, a dimensionless luminous efficiency of $1.4 \%$, and a beginning entry angle of \ang{15}, although we emphasize this is not a unique solution.

Note that the entry angle in the simulation is the entry angle relative to the surface of the Earth above the simulation start point, at a height of \SI{180}{\kilo \metre}. The change in the initial entry angle was caused by the curvature of the Earth as the ground distance between the beginning of the simulation and the first observed point was over \SI{800}{\kilo \metre}. Here a higher luminous efficiency had to be adopted compared to earlier because no fragmentation was included in the model. To reproduce the measured initial velocity at the meteor beginning height, we had to assume a velocity which was \SI{400}{\metre \per \second} higher at the beginning of the simulation at \SI{180}{\kilo \metre}, indicating that the semi-major axis was $\sim 0.4$ AU higher than the nominal value. 

Table \ref{tab:july21_radiant} lists the meteoroid's radiant and orbital elements, with and without the initial velocity correction. The meteoroid came from the antihelion source and was not associated with any known meteor shower. Its Tisserand's parameter with respect to Jupiter suggests it might have a Jupiter-family comet (JFC) origin. We believe that the ejection from its parent comet happened very recently as it was on a Jupiter crossing orbit, and such orbits have short dynamical lifetimes.

\begin{table*}[t] %floating table at the top of the page
	\caption{Radiant and orbital elements (in J2000) of the July 21, 2017 fragmenting meteor.}
	\label{tab:july21_radiant} % is used to refer this table in the text
	\centering % used for centering table
	
	\begin{tabular}{l r r l} % centered columns (5 columns)
	\hline\hline % inserts double horizontal lines
	
	Parameter                   & Nominal & $v_0 + \SI{400}{\metre \per \second}$ & Uncertainty* \\ % table heading
	\hline % inserts single horizontal line
	
	$\alpha_g$                  & 253.626 & 254.399 & \ang{0.017} \\
	$\delta_g$                  & -30.216 & -29.500 & \ang{0.103} \\
	$v_g$                       &  11.839 &  12.381 & \SI{0.007}{\kilo \metre \per \second}\\
	$\lambda_{\odot}$           & 118.482 & 118.482 & NA \\
	$L_g - \lambda_{\odot}$     & 137.292 & 137.883 & \ang{0.008} \\
	$B_g$                       &  -7.583 &  -6.796 & \ang{0.032} \\
	$a$                         &   3.199 &   3.588 & 0.005 AU \\
	$e$                         &   0.703 &   0.736 & 0.0005 \\
	$q$                         &   0.952 &   0.948 & 0.0001 AU\\
	$Q$                         &   5.447 &   6.229 & 0.010 AU\\
	$\omega$                    &  32.258 &  32.860 & \ang{0.010} \\
	$i$                         &   2.400 &   2.233 & \ang{0.033} \\
	$\pi$                       & 330.644 & 331.239 & \ang{0.094} \\
	$T_j$                       &   2.742 &   2.574 & 0.002 \\
	\hline %inserts single line
	\end{tabular} \\
	* uncertainties indicate measurement precision, not accuracy
\end{table*}

\subsubsection{Deceleration, strength, and mass distribution of fragments} \label{subsec:july21_event_dyn_press}

Narrow-field video data from Tavistock was reduced in \texttt{mirfit} by manually picking the centroids of all discernible fragments. The fragments were labeled from 1 to 12, according to their order of appearance. The celestial coordinates of each fragment were projected onto the trajectory estimated from wide-field observations. As there was no multi-station narrow-field data, only along-track positions of fragments could be precisely determined. Fragments 4 and 7 show a perpendicular offset from the trajectory (a sudden jump at the moment of fragmentation to a position parallel to other fragments), but only lower limits of transverse positions can be computed. Interestingly, these transversely offset fragments show no constant transverse velocity after fragmentation, indicating that they received and immediately lost momentum in the transverse direction. We are unsure what physical process caused this behaviour, but \cite{stokan2014transverse} give some possible explanations.

The exact moment of fragmentation could not be observed because the fragments were obscured by luminous dust. To extrapolate the motion of fragments shortly before they became observable, we fit a simplistic kinematic model proposed by \cite{jacchia1961precision} to the along-track distance of every fragment from the beginning of the meteor:

\begin{equation} \label{eq:jacchia_lag_func}
    D(t) = k + v_0 t + a_1 e^{a_2 t}
\end{equation}

\noindent where $t$ is the relative time since the beginning of the meteor, $v_0$ is the initial velocity of every fragment at $t = 0$, and $a_1$ and $a_2$ are deceleration coefficients. This model is plotted with all lag measurements throughout this work.

We assumed that the $v_0$ is equal to the meteor's initial velocity estimated from wide-field data. Next, we propagated the positions of fragments back in time using the model fits and identified when the positions intersected, which we took to be the time of a fragmentation event. Candidate fragmentation events were visually confirmed by inspecting the narrow-field video. We found that all fragments, except possibly 1 and 6, emerged from larger fragments. For fragments 1 and 6 it was not possible to visually confirm any prior points of fragmentation. As a singular point of fragmentation from which all fragments were born could not be determined, we believe that the observed fragments were created by progressive fragmentation; the 12 that were visible had strengths large enough not to fragment further for 1 second.

\begin{figure}
  \includegraphics[width=\linewidth]{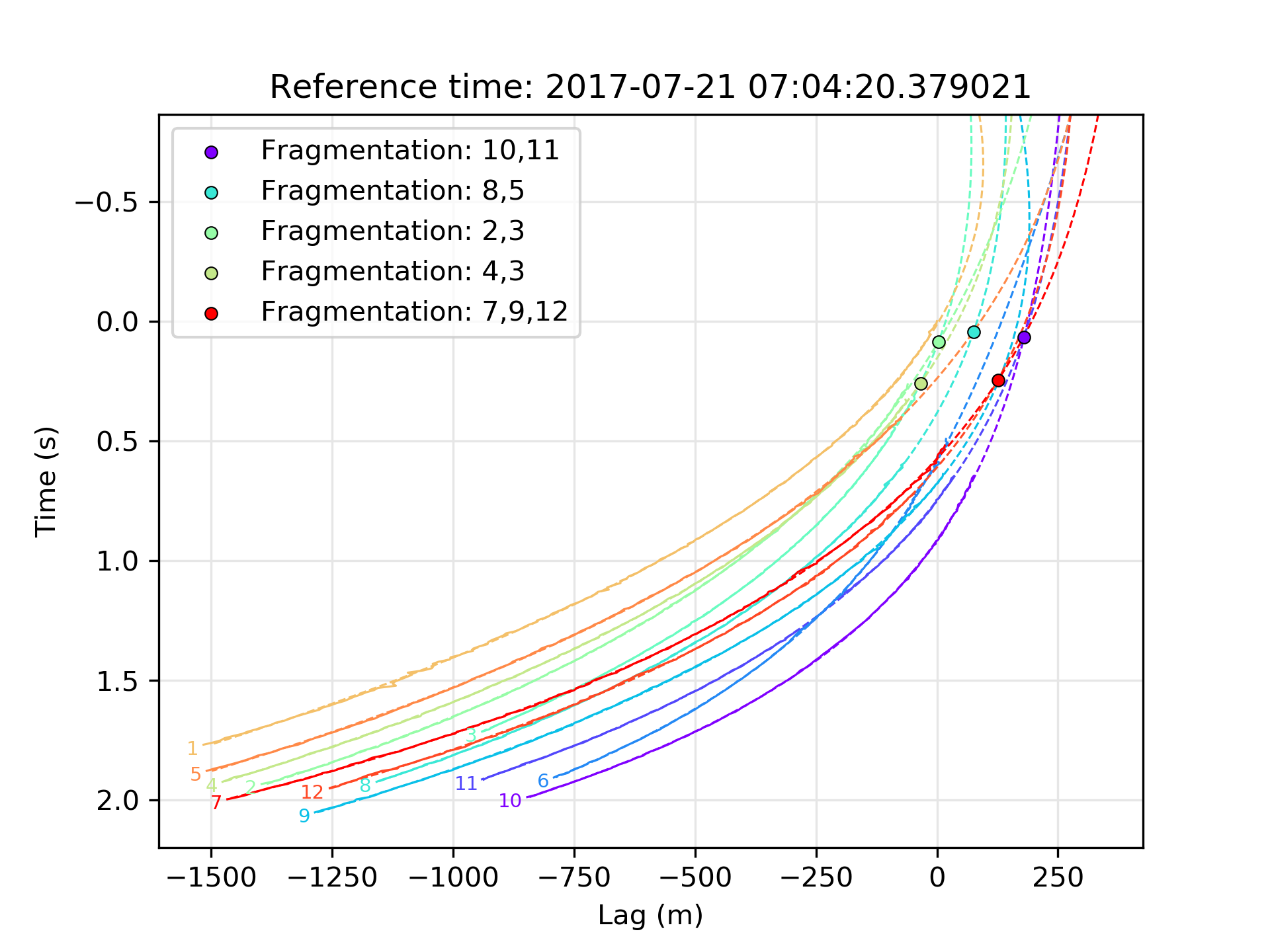} 
  \caption{Lags of individual fragments. Solid lines show the observed lag, the dashed lines show the extrapolated lag using the exponential function fit (equation \ref{eq:jacchia_lag_func}). Solid circles show the estimated points of fragmentation.}
  \label{fig:july21_narrow_lag_frags}
\end{figure}

Figure \ref{fig:july21_narrow_lag_frags} shows the lag of individual fragments, normalized to the first visible fragment which starts at \SI{0}{\second} and has a lag of \SI{0}{\metre}. Because all other fragments are in front of fragment 1, they all have a more positive lag. Fragment 10 was leading the ``fragment train''. The deceleration of fragments was not uniform, which caused fragments to overtake one another, indicative of an underlying mass distribution. 

Fragment 6 became visible in the middle of the ``fragment train'', but it overtook fragments 7, 12, 9, and 11 (in that order). It had the largest mass/area ratio and presumably had the largest mass, thus the smallest deceleration, indicating that there was no sorting by mass along the trajectory prior to fragmentation. A similar behaviour showing fragments overtaking one another has also been observed for fragmenting fireballs \citep{borovicka2003moravka}. All fragments were sufficiently separated in the transverse direction from one another as to not collide.

\begin{figure}
  \includegraphics[width=\linewidth]{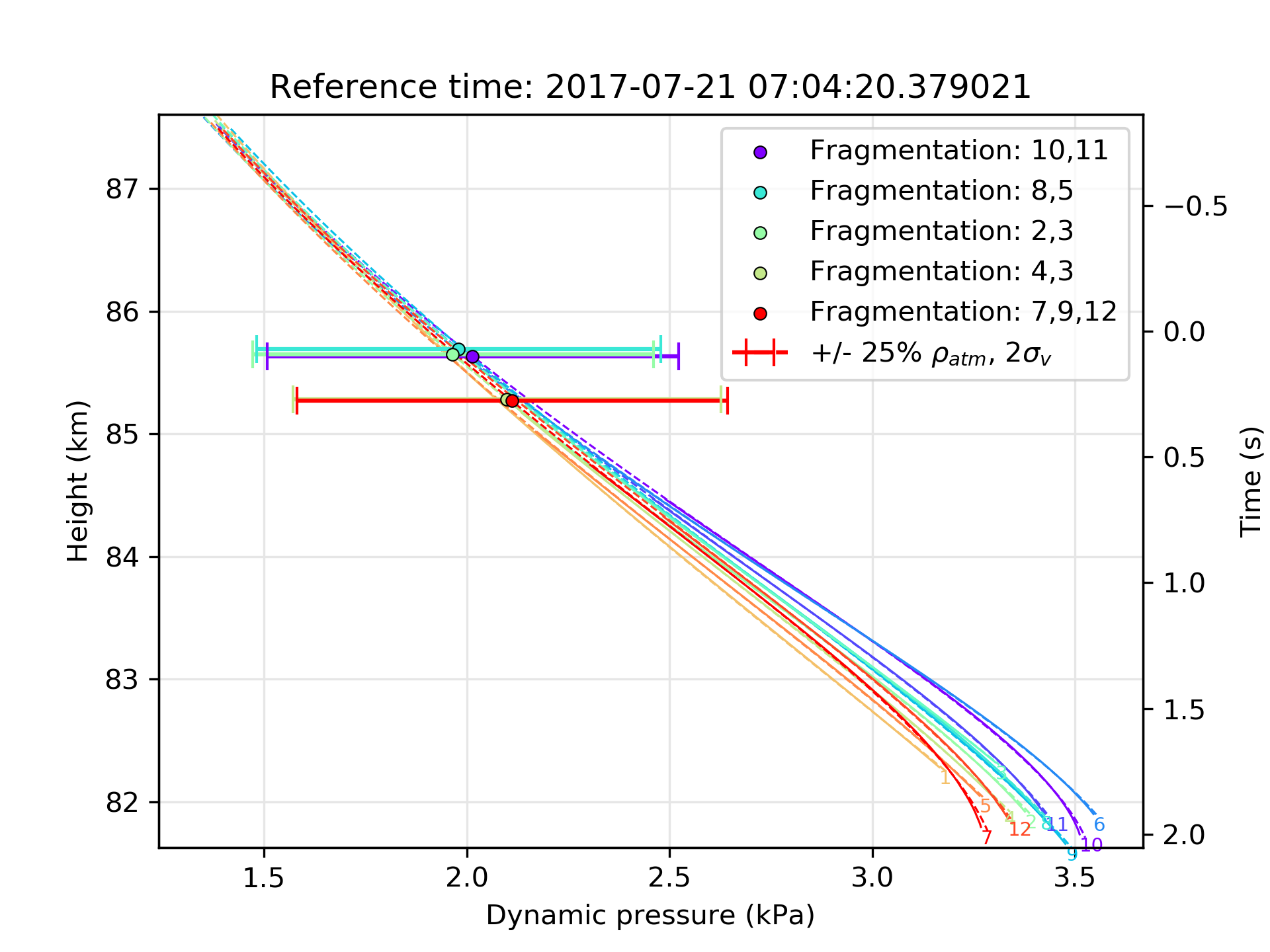} 
  \caption{Dynamic pressures of individual fragments. Solid circles mark the moments of fragmentation. The horizontal uncertainty bars show the spread in dynamic pressure at the fragmentation point due to the expected variance of the neutral atmosphere mass density.}
  \label{fig:july21_dyn_pressure}
\end{figure}

The dynamic pressure for every fragment is shown in Figure \ref{fig:july21_dyn_pressure}. The figure shows that the dynamic pressure at the moment of fragmentation was around \SI{2 \pm 0.5}{\kilo \pascal}. The fragments themselves started to disintegrate at a height of \SI{82.5}{\kilo \metre}, which corresponds to a dynamic pressure of around \SI{3}{\kilo \pascal}. This suggests that the upper limit of the compressive strength of more compact parts of fresh JFC material is in the range of \SIrange{2}{3}{\kilo \pascal}.

If the dust seen at the beginning of narrow-field observations is the eroding matrix in which these fragments were embedded  it would indicate that the upper limit of its overall global strength is \SI{1}{\kilo \pascal}, possibly on the order of several hundreds of pascals for cm-sized JFC meteoroids. Note that the erosion might have also been caused by thermal effects \citep{borovicka2007atmospheric}. In Figure \ref{fig:july21_lightcurve} we superimpose the estimated moments of fragmentation onto the wide-field light curve. The figure shows that the moments of fragmentation coincide with the first peak in the light curve, and that the fragments themselves started visibly disintegrating into their constituent grains at the beginning of the second peak.

These results are in accord with the in-situ measurements by the Philae lander \citep{biele2015landing} which estimated the surface strength of 67P to be \SIrange{1}{3}{\kilo \pascal}, a result similar to that found for individual particles by the COSIMA instrument \citep{hornung2016first}. Because all fragments appear to fragment at the same time and dynamic pressure, it appears that they have a similar drag coefficient, i.e. similar axial ratio or rapid rotation. We also note that the strength of the fragments themselves is only marginally larger than the strength at initial fragmentation. The fine separation of fragment strengths in this unusual case  was made possible by the low entry angle of the meteoroid which caused a very gradual increase in the dynamic pressure.

\subsubsection{Mass and size distribution of fragments}

We also attempted to measure the fragment mass distribution. The dynamic mass was computed by using the velocity from the exponential deceleration fit, due to which it was rapidly decreasing, indicating that the fragments themselves were eroding, although that did not become visibly obvious until they developed wakes at the end of luminous flight. The complete photometric mass of fragments was equally challenging to compute because the fragments were either very close to each other or enveloped in dust. 

\begin{figure}
  \includegraphics[width=\linewidth]{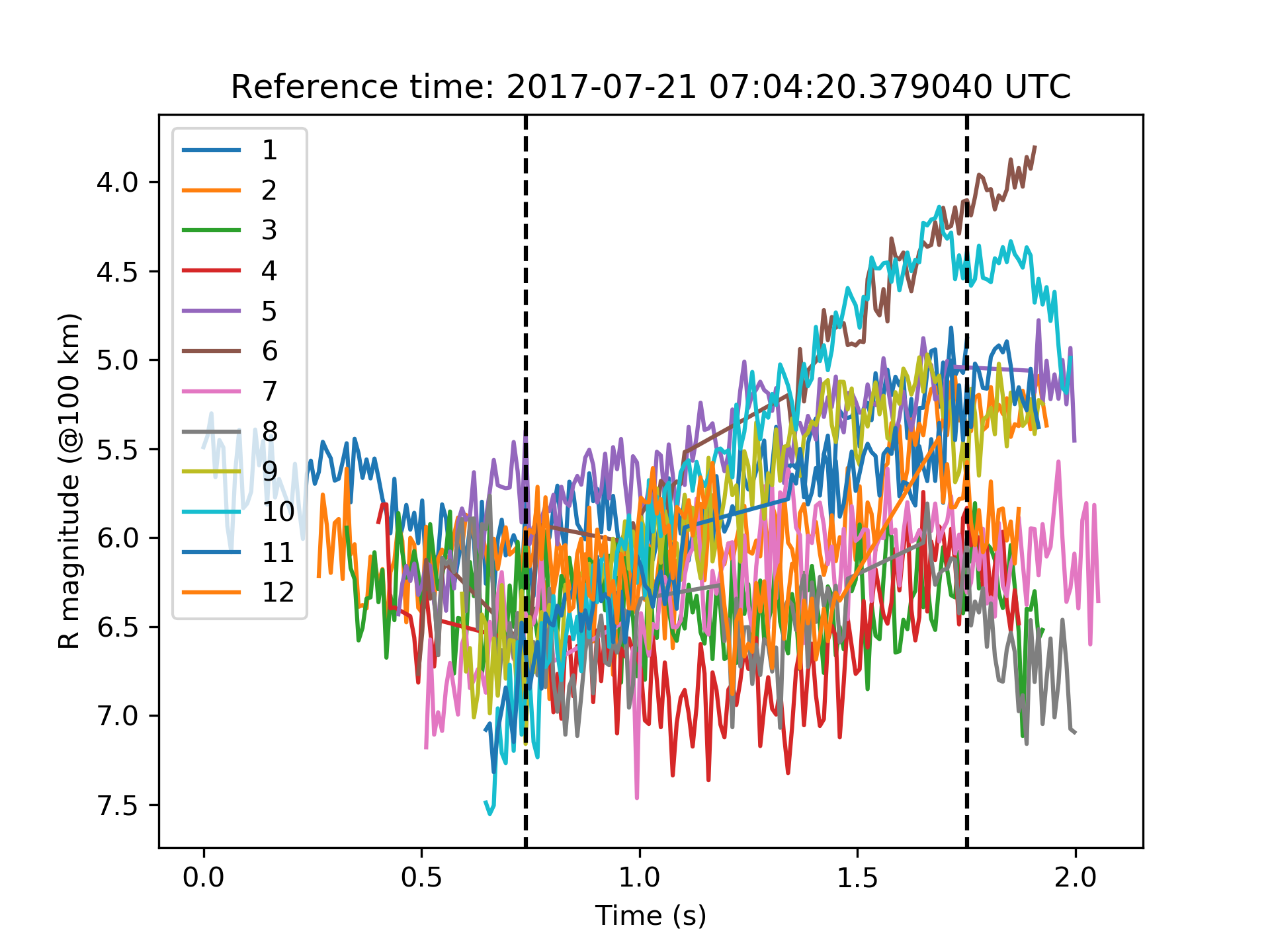} 
  \caption{Light curve of individual fragments. Vertical lines indicate the portion of time when all fragments were visible, and this was used to compute the partial fragment mass. Not all fragments were visible all the time, so we linearly interpolated the magnitudes in the gaps.}
  \label{fig:july21_frag_lightcurve}
\end{figure}

We were able to estimate lower limits to the photometric mass per fragment by measuring the brightness of fragments in one common interval when they were all clearly visible and distinct. Figure \ref{fig:july21_frag_lightcurve} shows the magnitude of every fragment; vertical lines mark the time range used for computing the mass, when all were visible. Table \ref{tab:july21_frag_masses} lists the computed masses using the same $\tau$ and $P_{0m}$ as used for the wide-field photometric mass, and diameters computed using a bulk density of $\rho = \SI{3000}{\kilogram \per \cubic \metre}$. Although these masses are half or less of their original value, their relative values to each other should be valid if we assume that they all started ablating at the same time and they ablated with a similar and constant ablation coefficient. 

We note that photometric masses of some fragments do not correspond to their dynamical behaviour. For example, fragment 5 was decelerating more than fragment 4, despite having three times larger photometric mass. This may indicate that these fragments had either different shapes, densities, or composition. Alternatively, some fragments may have been an unresolved group of smaller fragments.

\begin{figure}
  \includegraphics[width=\linewidth]{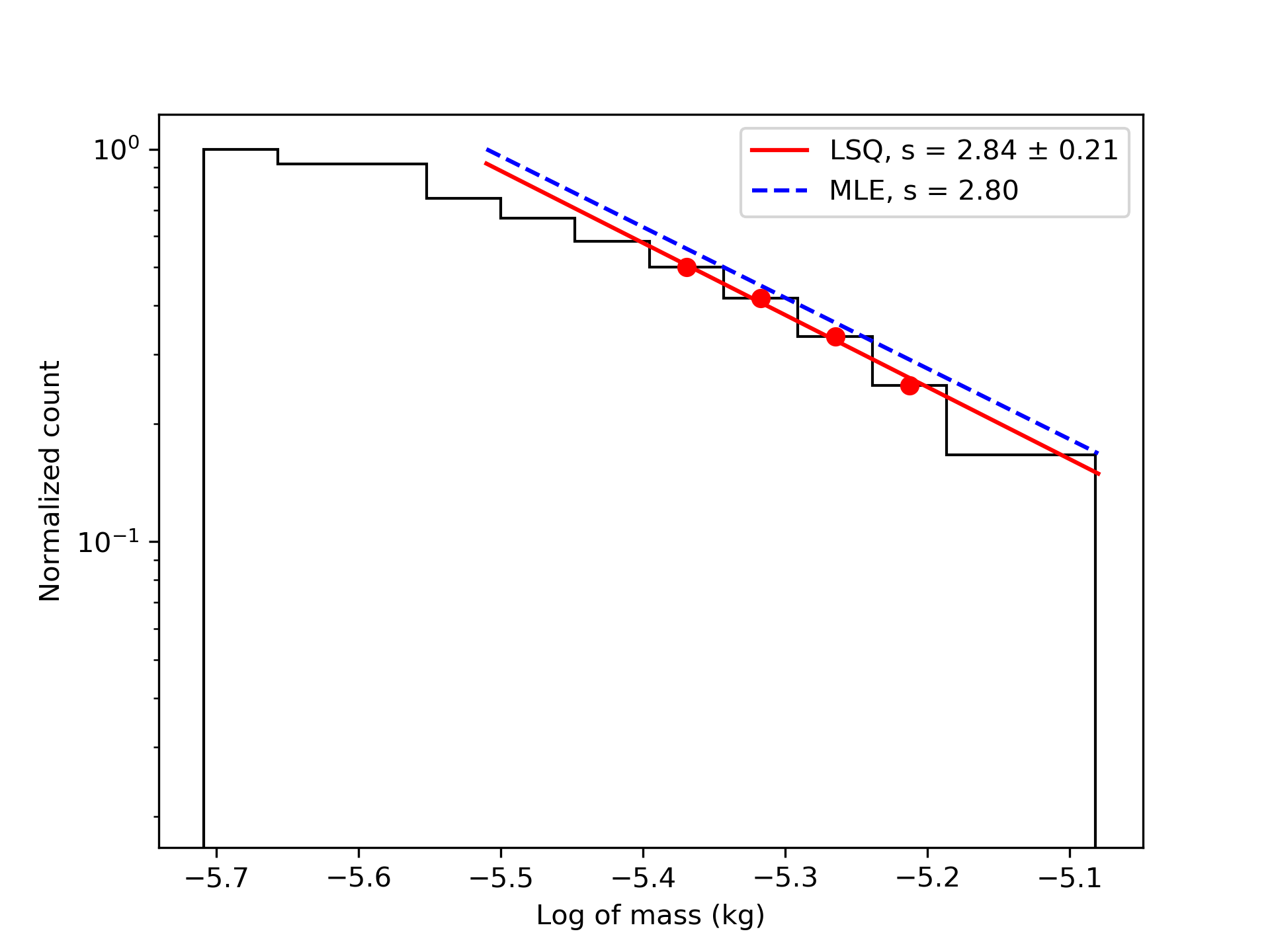} 
  \caption{Cumulative distribution of fragment masses of the July 21 event. The least squares fit (LSQ) on the selected range of masses (red dots) is shown in red, and the maximum likelihood estimation (MLE) fit is shown in blue.}
  \label{fig:july21_frag_mass_distribution}
\end{figure}

Figure \ref{fig:july21_frag_mass_distribution} shows the cumulative distribution of fragment masses. We estimated the mass index using two separate approaches. First, we performed a simplistic least squares (LSQ) line fit to the approximately linear part in the cumulative histogram. The measured mass index is $s = 2.84 \pm 0.21$, with the uncertainty only indicating the line fit uncertainty to those select points (red dots in the plot). This is not a robust approach of fitting power-law distributions to data \citep{clauset2009power}, so we performed a separate fit using maximum likelihood estimation (MLE) and obtained a value of $s = 2.80$, which is close to the line fit value. 

Following the procedure of \cite{alstott2014powerlaw}, we compared the goodness of fit between the power-law and the exponential distribution and found neither distribution is a significantly stronger fit (p-value = 0.29). The Kolmogotov-Smirnov D statistic \citep{ivezic2014statistics} was 0.19 for the power-law, and 0.13 for the exponential distribution, indicating that the latter is a slightly better fit. This indicates that either the fragment mass distribution was not a power-law, or that the power-law distribution quickly tapered off due to small number statistics. Thus, we believe that these $s$ values are lower limits.

Using the MLE approach, we also fitted a power-law to the distribution of fragment radii. The sizes changed with $\alpha_s = 6.4$, and the exponent is insensitive to the choice of bulk density. This is consistent with theoretical transformations where the size index exponent is equal to $\alpha_s = 3s - 2$ \citep[Appendix C in][]{vaubaillon2005new}. This is a very large exponent value compared to Rosetta measurements where they measured $\alpha_s = 1.8, s = 1.27$ for particles \SI{>150}{\micro \meter}, and $\alpha_s = 2.9, s = 1.63$ for particles in the \SIrange{30}{150}{\micro \metre} range \citep{merouane2016dust}. This steepness, and the fact that all observed fragments could not be tracked to a single fragmentation event, might indicate that the fragments we measured are daughter-fragments of progressive fragmentation, and that the observed fragments were not pre-existing in the meteoroid. Note that \cite{merouane2016dust} give a cumulative size index $\alpha_{sc}$ which relates to the size index $\alpha_s$ used here as $\alpha_s = \alpha_{sc} + 1$.

\begin{table*}[t] %floating table at the top of the page
	\caption{Estimated partial fragment masses and diameters of the July 21 event, sorted by increasing value. The diameters of spherical particles were computed using a bulk density of \SI{3000}{\kilogram \per \cubic \metre}.}
	\label{tab:july21_frag_masses} % is used to refer this table in the text
	\centering % used for centering table
	
	\begin{tabular}{l r r} % centered columns (5 columns)
	\hline\hline % inserts double horizontal lines
	
	Fragment No. & Mass (kg) & Diameter (mm) \\ % table heading
	\hline % inserts single horizontal line
	
	 4 & \num{1.96e-06} & 2.31 \\
	 8 & \num{2.53e-06} & 2.52 \\
	 3 & \num{2.62e-06} & 2.54 \\
	 7 & \num{3.06e-06} & 2.68 \\
	12 & \num{3.53e-06} & 2.81 \\
	 2 & \num{3.98e-06} & 2.92 \\
	11 & \num{4.10e-06} & 2.95 \\
	 9 & \num{5.02e-06} & 3.16 \\
	 1 & \num{5.33e-06} & 3.22 \\
	 5 & \num{5.99e-06} & 3.35 \\
	10 & \num{7.92e-06} & 3.68 \\
	 6 & \num{8.28e-06} & 3.73 \\

	\hline %inserts single line
	\end{tabular} \\
\end{table*}

\subsection{Identification and analysis of a larger population of fragmenting meteors} \label{subsec:strength_survey}

Having established and presented our analysis methodology in detail for this first case study, we expand our analysis to additional events. We identified 19 more events which showed gross fragmentation with measurable fragments, and list their details in Table \ref{tab:measured_meteors}. These events were not as favourable as the July 21 event because they had steeper entry angles and consequently shorter trajectories (resulting in fewer data points), and had fewer measurable fragments (usually only 2-3). The error in the initial height was not estimated because all moments of fragmentation were visually confirmed in the video, thus it was only perhaps off by one or two video frames. Again, we note that the initial velocity may be underestimated by several hundreds of meters per second. Also, the velocity uncertainty stated in the table is the relative measurement precision (the variance within our measurements), and not the real absolute accuracy. Detailed ablation modelling is needed to invert the latter \citep{vida2018modelling}. For events which had multiple fragmentations, we list their mean value.

The observed events span a range of velocities and orbital types, indicating that gross fragmentation is not restricted to any one orbital type \citep[a result also previously found by][]{subasinghe2016physical}. We emphasize that these events comprise only $5\%$ of all observed meteors; thus this sample should not be considered an unbiased survey of the entire meteoroid population. According to \cite{subasinghe2016physical}, 95\% of all meteors observed by the CAMO tracking system show no discernible fragments, and most (>85\%) have a distinct wake, likely caused by erosion of the meteoroid into constituent grains in the \SIrange{1}{100}{\micro \metre} size range, a process not triggered by mechanical failure \citep{borovicka2007atmospheric}.

\begin{figure}
  \includegraphics[width=\linewidth]{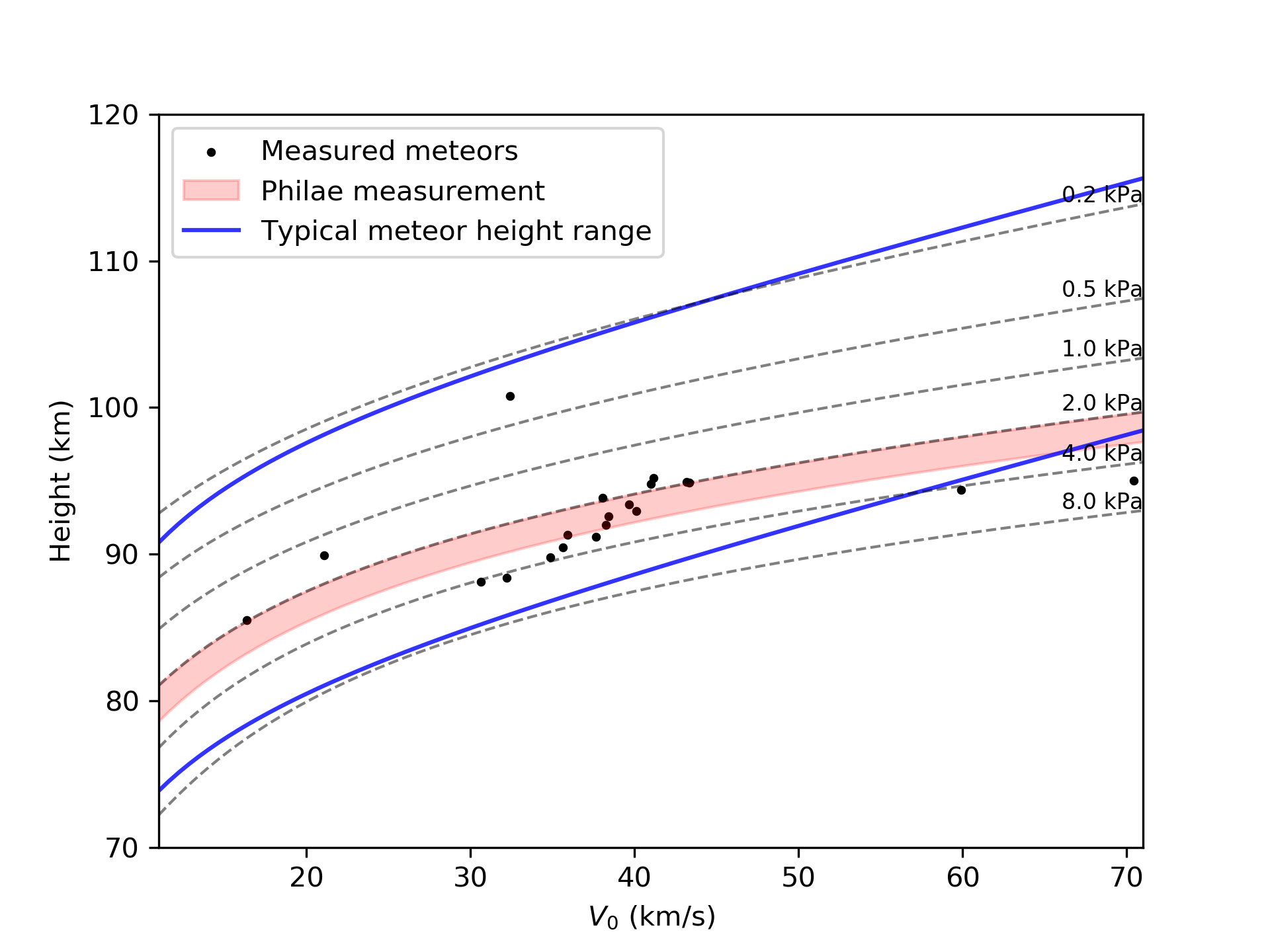} 
  \caption{Height vs. initial velocity of measured meteors. Dashed lines indicate contours of dynamic pressures at the given height and speed.}
  \label{fig:strengths_ht}
\end{figure}

Figure \ref{fig:strengths_ht} shows the measured compressive strengths based on the fragmentation height as function of initial meteoroid speed. The shaded zone represents the \SIrange{2}{3}{\kilo \pascal} strength range measured for 67P by the Philae lander \citep{biele2015landing}. The nominal strengths of most meteoroids lie in the range of \SIrange{1}{5}{\kilo \pascal}, in excellent accord with results reported by other authors discussed in sections \ref{subsec:strengths_intro} and \ref{subsec:strengths_atm}. 

\begin{figure}
  \includegraphics[width=\linewidth]{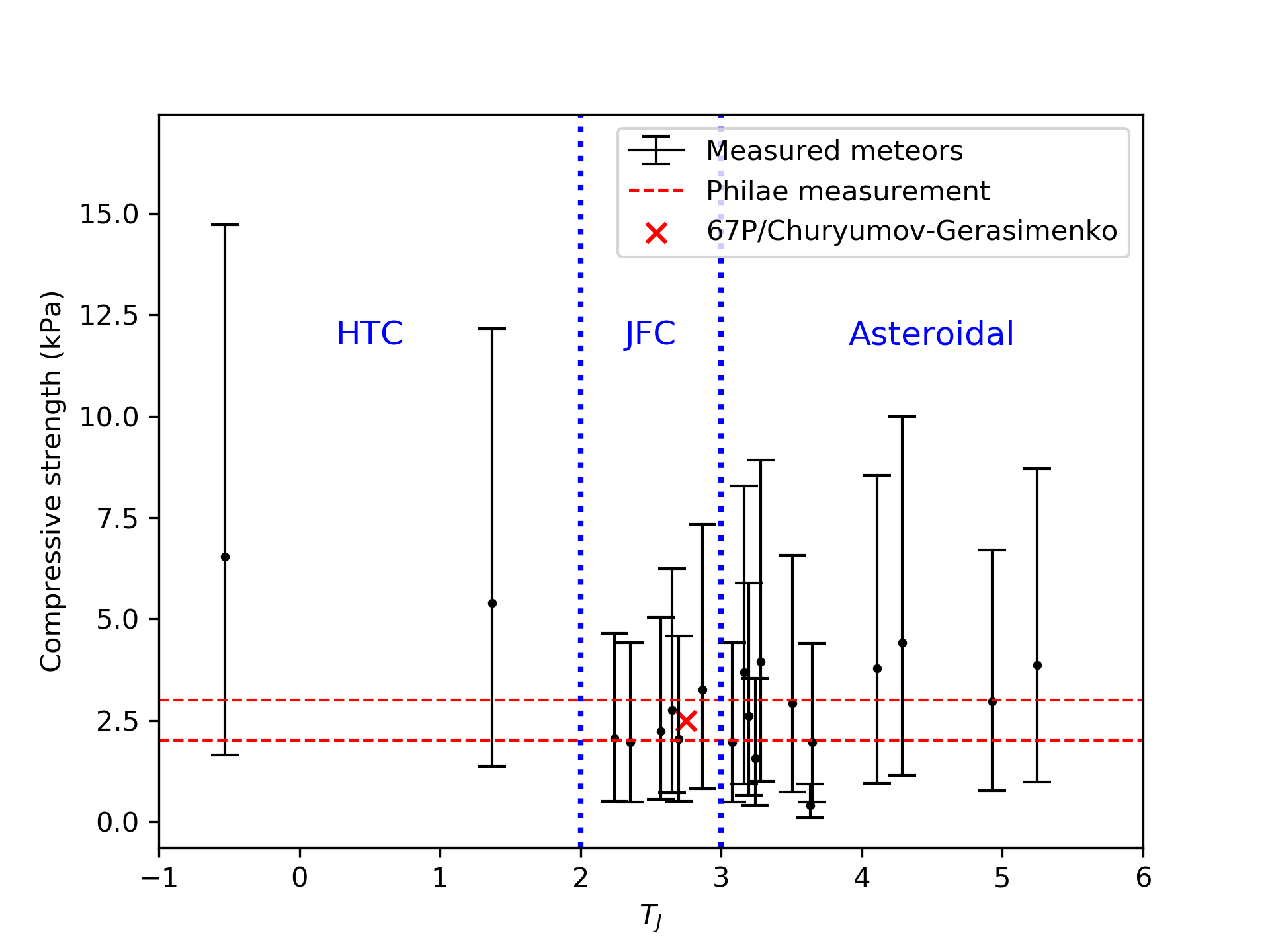} 
  \caption{Measured compressive strengths vs. Tisserand's parameter with respect to Jupiter. The error bars indicate the uncertainty range in dynamic pressure corresponding to an atmospheric mass density variance of $\pm 25\%$.}
  \label{fig:strengths_tj}
\end{figure}

Figure \ref{fig:strengths_tj} shows the measured compressive strengths versus the Tisserand's parameter with respect to Jupiter. There doesn't appear to be a trend in strength with orbital type. All measurements, except one, are within the measurement uncertainty of the Philae in-situ measured upper strength limits. Nevertheless, note that the statistical sample is small, and only the strengths of meteors with a particular morphology were measured. Also, we do not exclude the possibility of cross-contamination between objects on JCF and asteroidal orbits. Measured meteoroids on Halley-type orbits had higher strengths, but also higher uncertainties. As these meteors are very fast, determining the exact point of fragmentation is more difficult.

\begin{table*}[t] %floating table at the top of the page
	\caption{Radiants, orbits, and compressive strengths of measured meteoroids.}
	\label{tab:measured_meteors} % is used to refer this table in the text
	\centering % used for centering table
	\resizebox{\textwidth}{!}{ %% DELETE IN THE FINAL VERSION!
	\begin{tabular}{l r r r r r r r r r r r r r} % centered columns (5 columns)
	\hline\hline % inserts double horizontal lines
	
	Date and time (UTC) & $\lambda_{\odot}$ & $\alpha_g$ & $\delta_g$ & $v_g$ & $a$ & $e$ & $q$ & $\omega$ & $i$ & $\pi$ & $P_{dyn}$ \\
    & deg & deg & deg & \SI{}{\kilo \metre \per \second} & AU &  & AU & deg & deg & deg & kPa \\
\hline
2017-07-21 07:04:19 & 118.4821 & 253.866 & -30.059 & 11.995 & 3.299 & 0.7119 & 0.9505 & 32.44 & 2.37 & 330.83 & 2.03 \\
  &   & 0.019 & 0.084 & 0.009 & 0.007 & 0.0007 & 0.00007 & 0.01 & 0.03 & 0.01 & \\
2018-05-21 06:27:59 & 59.9221 & 318.411 & 16.967 & 58.748 & 2.298 & 0.5620 & 1.0065 & 190.02 & 124.07 & 249.94 & 5.40 \\
  &   & 0.210 & 0.730 & 0.064 & 0.176 & 0.0307 & 0.00101 & 1.06 & 1.02 & 1.06 & \\
2018-07-03 03:08:39 & 100.9155 & 311.625 & -6.031 & 41.826 & 2.256 & 0.9654 & 0.0780 & 331.58 & 46.57 & 72.50 & 2.23 \\
  &   & 0.022 & 0.107 & 0.033 & 0.010 & 0.0003 & 0.00068 & 0.14 & 0.34 & 0.14 & \\
2018-07-07 03:52:46 & 104.7580 & 311.131 & -4.558 & 33.827 & 1.396 & 0.8869 & 0.1580 & 322.83 & 28.51 & 67.60 & 3.79 \\
  &   & 0.044 & 0.296 & 0.013 & 0.005 & 0.0012 & 0.00124 & 0.14 & 0.56 & 0.14 & \\
2018-07-07 04:45:01 & 104.7926 & 312.437 & -4.826 & 38.075 & 1.834 & 0.9351 & 0.1191 & 325.79 & 35.65 & 70.59 & 2.61 \\
  &   & 0.039 & 0.085 & 0.006 & 0.004 & 0.0003 & 0.00075 & 0.12 & 0.18 & 0.12 & \\
2018-07-07 05:05:52 & 104.8064 & 312.009 & -4.428 & 36.768 & 1.663 & 0.9206 & 0.1321 & 324.59 & 33.89 & 69.41 & 2.92 \\
  &   & 0.031 & 0.103 & 0.006 & 0.003 & 0.0004 & 0.00075 & 0.11 & 0.20 & 0.11 & \\
2018-07-07 06:18:01 & 104.8541 & 320.981 & -26.329 & 39.515 & 1.882 & 0.9535 & 0.0875 & 150.75 & 36.55 & 75.60 & 1.96 \\
  &   & 0.019 & 0.073 & 0.015 & 0.006 & 0.0001 & 0.00051 & 0.10 & 0.19 & 0.10 & \\
2018-07-07 06:31:33 & 104.8631 & 275.387 & -19.770 & 17.874 & 2.277 & 0.6816 & 0.7250 & 252.88 & 1.97 & 357.88 & 1.56 \\
  &   & 0.097 & 0.178 & 0.098 & 0.021 & 0.0035 & 0.00177 & 0.15 & 0.09 & 0.15 & \\
2018-07-07 08:12:14 & 104.9298 & 316.019 & 2.403 & 34.192 & 1.124 & 0.8687 & 0.1476 & 327.02 & 44.33 & 71.95 & 2.97 \\
  &   & 0.091 & 0.129 & 0.061 & 0.004 & 0.0014 & 0.00181 & 0.25 & 0.17 & 0.25 & \\
2018-07-10 07:45:55 & 107.7733 & 323.502 & -28.420 & 36.462 & 1.593 & 0.9166 & 0.1328 & 144.88 & 34.33 & 72.64 & 1.95 \\
  &   & 0.031 & 0.049 & 0.019 & 0.003 & 0.0002 & 0.00033 & 0.05 & 0.13 & 0.05 & \\
2018-07-12 03:38:50 & 109.5174 & 312.697 & -2.902 & 36.609 & 2.137 & 0.9179 & 0.1754 & 316.79 & 31.93 & 66.31 & 3.26 \\
  &   & 0.022 & 0.096 & 0.016 & 0.006 & 0.0005 & 0.00073 & 0.09 & 0.15 & 0.09 & \\
2018-07-16 06:27:18 & 113.4454 & 318.755 & -1.198 & 38.760 & 2.273 & 0.9383 & 0.1403 & 321.31 & 38.05 & 74.77 & 2.76 \\
  &   & 0.230 & 0.533 & 0.090 & 0.045 & 0.0015 & 0.00467 & 0.78 & 1.14 & 0.78 & \\
2018-07-20 04:33:28 & 117.1871 & 320.915 & -1.886 & 35.987 & 1.890 & 0.9150 & 0.1607 & 319.63 & 29.20 & 76.82 & 3.68 \\
  &   & 0.045 & 0.240 & 0.012 & 0.008 & 0.0006 & 0.00092 & 0.12 & 0.50 & 0.12 & \\
2018-07-20 07:37:42 & 117.3092 & 327.577 & 5.895 & 33.230 & 1.071 & 0.8645 & 0.1451 & 328.06 & 41.88 & 85.38 & 3.86 \\
  &   & 0.055 & 0.163 & 0.049 & 0.003 & 0.0007 & 0.00108 & 0.17 & 0.36 & 0.17 & \\
2018-08-06 05:37:04 & 133.4761 & 346.893 & -11.180 & 41.877 & 2.571 & 0.9806 & 0.0499 & 157.01 & 23.95 & 110.48 & 2.06 \\
  &   & 0.042 & 0.116 & 0.006 & 0.019 & 0.0002 & 0.00071 & 0.19 & 0.37 & 0.19 & \\
2018-08-11 03:06:25 & 138.1700 & 329.209 & 29.510 & 28.636 & 1.414 & 0.6464 & 0.5002 & 286.86 & 42.50 & 65.04 & 4.42 \\
  &   & 0.094 & 0.303 & 0.086 & 0.008 & 0.0009 & 0.00383 & 0.53 & 0.31 & 0.53 & \\
2018-08-15 02:45:29 & 141.9972 & 347.331 & 1.497 & 39.616 & 2.582 & 0.9650 & 0.0903 & 328.72 & 20.83 & 110.73 & 1.96 \\
  &   & 0.031 & 0.083 & 0.008 & 0.005 & 0.0001 & 0.00033 & 0.06 & 0.26 & 0.06 & \\
2018-08-15 06:15:57 & 142.1377 & 344.188 & -11.273 & 30.427 & 1.698 & 0.8581 & 0.2409 & 130.59 & 6.49 & 92.71 & 0.41 \\
  &   & 0.061 & 0.168 & 0.012 & 0.004 & 0.0001 & 0.00058 & 0.09 & 0.27 & 0.09 & \\
2018-09-17 04:13:18 & 173.9824 & 8.893 & 5.885 & 30.127 & 1.935 & 0.8591 & 0.2727 & 305.20 & 2.69 & 119.22 & 3.95 \\
  &   & 0.095 & 0.240 & 0.037 & 0.010 & 0.0006 & 0.00094 & 0.14 & 0.36 & 0.14 & \\
2018-09-19 08:55:05 & 176.1250 & 87.333 & 13.044 & 69.437 & 7.693 & 0.8695 & 1.0040 & 357.02 & 162.08 & 353.15 & 6.53 \\
  &   & 0.076 & 0.189 & 0.047 & 0.397 & 0.0063 & 0.00010 & 0.25 & 0.32 & 0.25 & \\
	\hline %inserts single line
	\end{tabular}} \\
	* uncertainties of trajectory parameters indicate measurement precision, not accuracy
\end{table*}

\section{Conclusions}

We have summarized hardware and software reduction details of the upgraded CAMO mirror tracking system, compared to the original instrument \citep{weryk2013camo}. The current CAMO system achieves an effective astrometric precision of 1 arc second, and a temporal resolution of 10 ms. 

Using three representative types of meteors observed with CAMO, we have shown that the resolved trail morphology is the limiting factor in precision and ultimately, the obtainable trajectory accuracy. In ideal conditions, CAMO achieves trajectory fit precision of \SI{< 1}{\metre} and initial velocity measurement precision on the order of \SI{1}{\metre \per \second}. Both the radiant and speed of heavily fragmenting and eroding meteors have an order of magnitude higher uncertainty than meteors with short wakes. For highly fragmenting meteors, the radiant precision is similar to what can be achieved with moderate field of view non-tracking systems.

We used direct observations at the instant of gross meteoroid fragmentation to measure the compressive strengths of meteoroids. We used the aerodynamic ram pressure exerted on the meteoroid at the moment of fragmentation as a proxy for the compressive strength. A case study of an unusually long and shallow entry event on July 21, 2017, where narrow-field video showed the exact moment of fragmentation, resulted in 12 distinct fragments whose positions were tracked. A very shallow entry angle of \ang{8} enabled precise determination of the moments of fragmentation, and consequently precise strength measurements. The meteoroid started eroding at dynamic pressures below \SI{1}{\kilo \pascal} - this was not observed directly, but it was deduced from the long wake visible at the beginning of narrow-field tracking. We note that the cause of erosion might be thermal and not due to mechanical failure.

Next, the meteoroid visibly fragmented at \SI{2}{\kilo \pascal}, and the fragments themselves disintegrated at \SI{3}{\kilo \pascal}. We estimated a fragment mass index of $s = 2.8$ but  believe this to be a lower limit. This value is much larger than that derived from in-situ measurements by Rosetta of comet 67P's dust, and also larger than the mass indices of major meteor showers. This may indicate that the observed fragments were not pre-existing in the meteoroid matrix, but created by progressive fragmentation.

Nineteen more meteors showing visible fragments after gross fragmentation were used to survey compressive strengths. The majority had compressive strengths between \SIrange{1}{4}{\kilo \pascal} which were not correlated with orbital type. These measurements are in excellent accordance with the in-situ measurements of comet 67P by the Philae lander (where the maximum strength was between \SIrange{2}{3}{\kilo \pascal}) and the Rosetta COSIMA instrument (strength on the order of several kPa), as well as other theoretical and experimental work summarized in Section \ref{subsec:strengths_intro}.

The overall measurement uncertainty of compressive strengths was about $\pm 25\%$ due to the uncertainty in the atmosphere mass density. We note that only 5\% of all meteors observed by CAMO show gross fragmentation, thus these measurements do not represent all meteoroids, although both cometary and asteroidal orbits are represented in this sample.

Having developed and validated these methods for the analysis of high temporal and spatial resolution meteors observed with CAMO, in the future we aim to accurately measure the orbits of select meteor showers and use the high-precision constraints set by CAMO to improve meteor shower prediction models. The focus of our future work will be on those meteor showers caused by recently ejected meteoroids whose dispersion is solely caused by their ejection velocity from the parent body, as gravitational and non-gravitational forces do not have time to disperse the stream on such short timescales.

\section{Acknowledgements}
We thank the anonymous reviewers for their careful reading of our manuscript and their insightful suggestions. We thank Beau Bierhaus and Dr. Holly Capelo for useful discussions about comet strengths. We also thank Jason Gill for data management and format
conversion, and Zbigniew Krzeminski for hardware support.

This work was supported in part by the NASA Meteoroid Environment Office under cooperative agreement 80NSSC18M0046. PGB also acknowledges funding support from the Natural Sciences and Engineering Research council of Canada and the Canada Research Chairs program. JPM acknowledges support from the Chief of Naval Research and a grant of computing time from the US Department of Defense High Performance Computing Modernization Program. 

% Import the bibliography file bibliography.bib
\bibliography{bibliography}

\appendix

\section{Plate formats}

\subsection{AST plate} \label{appendix:ast_plate}

The AST (ASTrometry) plate is a type of plate mapping developed in \cite{weryk2012simultaneous} for use with the ASGARD system. It maps any cartesian (x, y) pair (e.g. image or mirror coordinates) into celestial horizontal coordinates ($\theta$, $\varphi$), where $\theta$ is the zenith distance and $\varphi$ is the azimuth (+N of due E). To avoid the discontinuity at the azimuth branch cut (where $\varphi = \pm 180$), the angles on a hemisphere (assuming that only $\theta < 90 \degree$ angles are observable) are projected onto a plane. The vertical axis of the projection, defined by angles ($\theta_0$, $\varphi_0$), is chosen to correspond close to the image centre. Thus, we can define the rotation matrix of the projection:
        
\begin{equation}
    M = 
    \begin{bmatrix}
        -\sin \varphi_0 & -\cos \theta_0 \cos \varphi_0 & \sin \theta_0 \cos \varphi_0 \\
         \cos \varphi_0 & -\cos \theta_0 \sin \varphi_0 & \sin \theta_0 \sin \varphi_0 \\
                      0 &                 \sin \theta_0 &                \cos \theta_0
    \end{bmatrix}
\end{equation}
        
The columns of matrix $M$ define an orthogonal basis set. Star positions ($\theta$, $\varphi$) are rotated using $M'$ (the inverse of $M$) to obtain ($\beta$, $\gamma$) pairs:
        
\begin{equation}
    \begin{bmatrix}
        \sin \beta \cos \gamma \\
        \sin \beta \sin \gamma \\
        \cos \beta
    \end{bmatrix}
    = M'
    \begin{bmatrix}
        \sin \theta \cos \varphi \\
        \sin \theta \sin \varphi \\
        \cos \theta
    \end{bmatrix}
\end{equation}

which are relative to ($\theta_0$, $\varphi_0$). The positions are then projected onto the $p, q$ plane:

\begin{equation}
\begin{aligned}
    p = \sin \beta \cos \gamma \\
    q = \sin \beta \sin \gamma
\end{aligned}
\end{equation}

The ($x$, $y$) image centroids are then fitted to the ($p$, $q$) values for each star using third order polynomials:

\begin{equation}
\begin{aligned}
    p = a_0 + a_1 x + a_2 x^2 + a_3 x^3 + a_4 y + a_5 y^2 \\ + a_6 y^3 + a_7 xy + a_8 x^2 y + a_9 x y^2 \\
    q = b_0 + b_1 x + b_2 x^2 + b_3 x^3 + b_4 y + b_5 y^2 \\ + b_6 y^3 + b_7 xy + b_8 x^2 y + b_9 x y^2
\end{aligned}
\end{equation}

The reverse mapping polynomials are fit separately, enabling the conversion from ($p$, $q$) to ($x$, $y$). As discussed in \cite{weryk2013camo}, this method is advantageous in comparison to a typical gnomonic projection for larger fields of view, by producing smaller fit residuals.

\subsection{AFF plate} \label{appendix:aff_plate}

The AFF plate represents an affine transform, which is a combination of translation, scaling, rotation and mirroring, described by the following equation:

\begin{equation}
    \begin{bmatrix}
    x' \\
    y'
    \end{bmatrix}
    = 
    \begin{bmatrix}
    M_{11} & M {12} & M_{13} \\
    M_{21} & M_{22} & M_{23}
    \end{bmatrix}
    \begin{bmatrix}
    x \\
    y \\
    1
    \end{bmatrix}
\end{equation}

The $M$ coefficients are the fit parameters. In this implementation, $x'$ and $y'$ are orthogonal, meaning that there is no shearing.

\end{document}